\newif\ifdraft
\newif\iffull
\newif\ifcomment
\newif\iflatexdiff
\newif\ifbibtex
\newif\ifpreprint
\latexdifftrue    
\fulltrue        
\bibtextrue       
\preprinttrue     
\def\dvers{v0.8}
\def\snntitle{$\snn$}
\ifpreprint
\def\snntitle{$\snnbf$}
\fi
\def\dtitle{Measurement of jet radial profiles\\ in Pb--Pb collisions at \snntitle = 2.76 TeV} 
\def\stitle{Jet radial profiles in Pb--Pb collisions} 
\ifpreprint
\documentclass[ALICE,manyauthors,12pt]{cernphprep}
\usepackage[comma,square,numbers,sort&compress]{natbib}
\else
\documentclass[final,3p,12pt]{elsarticle}
\biboptions{comma,square,numbers,sort&compress}
\fi
\usepackage{hyperref}
\usepackage{graphicx}  
\usepackage{dcolumn}   
\usepackage{bm}        
\usepackage{amssymb}   
\usepackage{amsfonts}
\usepackage{graphics}
\usepackage{grffile}   
\usepackage{epsfig}
\usepackage{units}
\usepackage{comment}
\usepackage[usenames]{color}
\usepackage[normalem]{ulem} 
\usepackage{color}
\usepackage[utf8]{inputenc}
\usepackage[T1]{fontenc}
\usepackage{subfigure}
\iflatexdiff
\RequirePackage{color}\definecolor{RED}{rgb}{1,0,0}\definecolor{BLUE}{rgb}{0,0,1}

\fi
\ifpreprint
\else
 
\fi

\newcommand{\com}[1]       {\relax}

\newcommand{\pp}           {pp}

\newcommand{\PbPb}         {\mbox{Pb--Pb}}

\newcommand{\pt}           {\ensuremath{p_{\mathrm{T}}}{ }}

\newcommand{\snn}          {\ensuremath{\sqrt{s_{\mathrm{NN}}}}}
\newcommand{\snnbf}        {\ensuremath{\mathbf{{\sqrt{\textit{s}_{NN}}}}}}

\newcommand{\Ref}[1]       {Ref.~\cite{#1}}
\newcommand{\Refs}[1]      {Refs.~\cite{#1}}

\newcommand{\red}[1]       {\textcolor{red}{#1}}

\newcommand{\warn}[1]      {{\small\textbf{\red{(!}\footnote{\textbf{\red{(!)}}~#1}\red{)}}}\marginpar{\textbf{\red{---}}}}

\iflatexdiff

\renewcommand{\xout}[1]    {\textcolor{red}{\sout{#1}}}
 
\else

\renewcommand{\xout}[1]    {}
\fi

\newcommand{\pT}           {\ensuremath{p_{\mathrm{T}}}}

\newcommand{\FastJet}      {\ensuremath{\mathrm{FastJet}}}
\newcommand{\kT}           {\ensuremath{k_{\mathrm{T}}}}
\newcommand{\jetraw}       {\ensuremath{{p}_{\mathrm{T,\,jet}}^{\mathrm{raw}}}}

\newcommand{\GeVc}         {GeV/$c$}

\newcommand{\Ajet}         {\ensuremath{{A}_{\mathrm{jet}}}}

\newcommand{\eq}[2]{\begin{equation}\label{#1} #2 \end{equation}}

\graphicspath{{./img/}}
\ifdraft
\usepackage{lineno}
\modulolinenumbers[5]
\usepackage{fancyhdr}
\else
\renewcommand{\warn}[1]{}

\fi

\begin{document}
\newlength{\figlen}
\setlength{\figlen}{\linewidth}
\ifpreprint
\setlength{\figlen}{0.95\linewidth}
\begin{titlepage}
\PHyear{2019}
\PHnumber{079}                  
\PHdate{30 April}             
\title{\dtitle}
\ShortTitle{\stitle}
\Collaboration{ALICE Collaboration%
         \thanks{See Appendix~\ref{app:collab} for the list of collaboration members}}
\ShortAuthor{ALICE Collaboration} 
\ifdraft
\begin{center}
\today\\ \color{red}DRAFT \dvers\ \hspace{0.3cm} \$Revision: 5255 $\color{white}:$\$\color{black}\vspace{0.3cm}
\end{center}
\fi
\else
\begin{frontmatter}
\title{\dtitle}
\iffull
\input{authors-plb.tex}
\else
\ifdraft
\author{ALICE Collaboration \\ \vspace{0.3cm} 
\today\\ \color{red}DRAFT \dvers\ \hspace{0.3cm} \$Revision: 5255 $\color{white}:$\$\color{black}}
\else
\author{ALICE Collaboration}
\fi
\fi
\fi
\begin{abstract}
The jet radial structure and particle transverse momentum ($\pT$) composition within jets are presented in centrality-selected \PbPb\ collisions at $\snn = 2.76$~TeV.
Track-based jets, which are also called charged jets, were reconstructed with a resolution parameter of $R=0.3$ at midrapidity $|\eta_\mathrm{ch\,jet}|<0.6$ for transverse momenta $p_\mathrm{T,\,ch\,jet}=30$--$120$~\GeVc.
Jet--hadron correlations in relative azimuth and pseudorapidity space~($\Delta\varphi$, $\Delta\eta$) are measured to study the distribution of the associated particles around the jet axis for different $p_\mathrm{T, assoc}$-ranges between 1 and 20~\GeVc.
The data in \PbPb\ collisions are compared to reference distributions for pp collisions, obtained using embedded PYTHIA simulations.
The number of high-$\pt$ associate particles~($4<p_\mathrm{T, assoc}<20$~\GeVc) in \PbPb\ collisions is found to be suppressed compared to the reference by 30 to 10\%, depending on centrality. 
The radial particle distribution relative to the jet axis shows a moderate modification in \PbPb\ collisions with respect to PYTHIA.
High-$\pT$ associate particles are slightly more collimated in \PbPb\ collisions compared to the reference, while low-$\pT$ associate particle tend to be broadened.
The results, which are presented for the first time down to $p_\mathrm{T,\,ch\,jet}=30$~\GeVc\ in \PbPb\ collisions, are compatible with both previous jet--hadron-related
measurements from the CMS Collaboration and jet shape measurements from the ALICE Collaboration at higher $\pt$, and add further support for the established picture of in-medium parton energy loss.

\ifdraft 
\ifpreprint
\end{abstract}
\end{titlepage}
\setcounter{page}{2}
\else
\end{abstract}
\end{frontmatter}
\newpage
\fi
\fi
\ifdraft
\thispagestyle{fancyplain}
\else
\end{abstract}
\ifpreprint
\end{titlepage}
\setcounter{page}{2}
\else
\end{frontmatter}
\fi
\fi
\section{Introduction}
\label{sec:intro}
At energy densities above approximately $0.5$~GeV/fm$^3$ and temperatures above approximately $160$ MeV~\cite{Bhattacharya:2014ara},
Quantum Chromodynamics~(QCD) calculations on the lattice predict the existence of a phase transition from normal nuclear matter to a new state of matter called the Quark--Gluon Plasma~(QGP), where the partonic 
constituents, quarks and gluons, are no longer confined in hadrons.
There is compelling evidence from observations reported by experiments at
the Relativistic Heavy Ion Collider~(RHIC)~\cite{Arsene20051,Back200528,Adcox2005184,Adams2005102} and at the Large Hadron 
Collider~(LHC)~\cite{Aamodt:2010pb,Aamodt:2010cz,Chatrchyan:2011pb,Aamodt:2011mr,Aamodt:2010pa,ATLAS:2011ah,Chatrchyan:2012wg,ALICE:2011ab,
Aad:2013xma,Chatrchyan:2013kba,Aamodt:2010jd,Chatrchyan:2011sx} that the QGP is created in nuclear collisions at high collisions energies.

A unique way to characterize the properties of the QGP is to utilize jets as a probe of the medium.
Hard scatterings are expected to occur early in the collision evolution, producing high transverse momentum~(\pT) partons 
that propagate through the expanding medium and eventually fragment into jets of hadrons. 
High-$\pT$ partons lose energy in interactions with the medium due to elastic scattering and induced gluon radiation~\cite{Gyulassy:1990ye,Baier:1994bd}.
Besides a reduction of the jet energy, this can result in a broadening of the transverse jet profile and a softening of the fragmentation~\cite{Salgado:2003rv}.

Jet quenching has been observed at RHIC~\cite{Adcox:2001jp,Adler:2002tq,Adler:2002xw,Adcox:2002pe,Adler:2003qi,Adams:2003kv,
Adams:2003im,Back:2003qr,Arsene:2003yk,Adams:2006yt,Adare:2006nr,Adare:2008cg,Adamczyk:2013jei,Adamczyk:2017yhe} and at the LHC~\cite{Aamodt:2010jd,Aad:2010bu,Chatrchyan:2011sx,Chatrchyan:2011pb,
Aamodt:2011vg,CMS:2012aa,Chatrchyan:2012nia,Chatrchyan:2012gw,Chatrchyan:2012gt,Aad:2012vca,Chatrchyan:2013exa,Chatrchyan:2013kwa,
Chatrchyan:2014ava,Aad:2014wha,Aad:2014bxa,Adam:2015doa},
e.g.\ via inclusive yield and correlation measurements of high-$\pT$ hadrons and reconstructed jets.
These measurements provide insights into the mechanisms of parton energy loss in the medium and eventually into the medium itself.

More differential measurements of the jet modification in a medium, i.e.\ measurements of modifications of jet angular profile and particle composition,
can provide complementary information to observables that focus on the overall yield change like nuclear modification factors.
Measurements of correlated associated particle production relative to jets or high-$\pT$ particles allow for a detailed measurement of the redistribution of quenched energy around the jet.
An excess of low-$\pT$ particles in and around the jet up to large distances, as well as
a suppression of high-$\pT$ particles, have been reported~\cite{Chatrchyan:2011sx,Adam:2016a,Adam:2016b,CMS:2016a}.
Two-particle correlations and jet--hadron correlations show an angular broadening of low-$\pT$ particles below 3~\GeVc\ in heavy-ion collisions with respect to \pp\ collisions~\cite{CMS:2016a}.
For low-$\pT$ two-particle correlations, measurements also indicate an asymmetry in the shape of the near-side jet peaks: they are broader in $\Delta\eta$ compared to $\Delta\varphi$~\cite{Adam:2016a,Adam:2016b}. The variables $\Delta\eta$ and $\Delta\varphi$ are the distance in pseudorapidity $\eta$ and azimuth $\varphi$ relative to the near-side jet.
At the same time, measurements of the radial moment of jets point to a general collimation of jets in \PbPb\ collisions~\cite{Acharya:2018uvf}.

Using jets instead of high-$\pT$ particles as a reference (trigger) to study angular correlations---as done in this analysis---should have the advantage that jet properties better reflect the initial parton energy.
This analysis extends the study of jet--hadron correlations into a regime of
low track-based jet $p_\mathrm{T,\,ch\,jet}$ not yet explored with these techniques at the LHC.

In this paper, we study the correlation of charged particles (associates) with the direction of reconstructed track-based jets (triggers) in the $\Delta\varphi$-$\Delta\eta$ plane in the same event.
The jets are reconstructed using charged particles above a certain transverse momentum $p_\mathrm{T,const}$.
The analysis focuses on two aspects of the modification of jets within the medium created in \PbPb\ collisions compared to a PYTHIA~\cite{Sjostrand:2006za} reference.
First, the overall modification of the associated particle yield and its jet-energy dependence is studied.
Second, the modification of the radial distribution of associated particles with respect to the jet axis is studied by comparing the \PbPb\ results to the PYTHIA reference.
Both aspects are analyzed in detail for several jet transverse momenta $p_\mathrm{T,\,ch\,jet}$ and low and high \pT\ of associated charged particles.
PYTHIA is used as vacuum baseline, because the size of the pp dataset at $\sqrt{s} = 2.76$~TeV is insufficient for this analysis.

The paper is structured as follows.
In Sec.~\ref{sec:setup}, details on the detector and general data reconstruction will be given.
The correlation analysis, which serves as basis for this paper, is presented in Sec.~\ref{sec:correlationanalysis}.
Subsequently, jet reconstruction will be described in Sec.~\ref{sec:reconstruction}, followed by a discussion on the embedded PYTHIA reference in Sec.~\ref{sec:embeddedBaseline}.
Before the results will be presented in Sec.~\ref{sec:results}, the observables are introduced in Sec.~\ref{sec:observables} and systematic uncertainties are discussed in Sec.~\ref{sec:systematics}. 
A summary concludes the paper in Sec.~\ref{sec:summary}.

\section{Experimental setup}
\label{sec:setup}
For a complete description of the ALICE detector and its performance see Refs.~\cite{Aamodt:2008zz} and~\cite{Abelev:2014ffa}, respectively.

The data were recorded in 2011 for \PbPb\ collisions at $\snn=2.76$ TeV using a set of centrality triggers based on the hit multiplicity measured by the V0 detector, 
which consists of segmented scintillators covering the full azimuth over $2.8<\eta<5.1$ (V0A) and $-3.7<\eta<-1.7$ (V0C).
Events were selected with V0 multiplicities corresponding to the $0$--$50$\% most central events using the centrality determination as described in \Ref{Abelev:2013qoq}.
The accepted events, reconstructed as described in \Ref{Abelev:2012hxa}, were required to have a primary reconstructed vertex within 
$\com{|{z}_{\rm vertex}|<}10$~cm of the center of the detector along the beam axis.
For this analysis, a total of $12$M events were used. 

The analysis presented here relies mainly on the central ALICE tracking systems, 
which are located inside a large solenoidal magnet with a field strength of $0.5$~T. 
They consist of the Inner Tracking System (ITS), a high-precision six-layer cylindrical silicon detector system with the inner layer at a radius of $3.9$~cm and the outer layer at $43$~cm
from the beam axis, and the Time Projection Chamber (TPC) with a radial extent of $85$--$247$~cm, which provides up to 159 independent space points per track.

To ensure a good track-momentum resolution for jet reconstruction, all reconstructed tracks were required to have at least three hits in the ITS.
For tracks without any hit in the Silicon Pixel Detector (SPD), which provides the two innermost layers of the ITS, the location of the primary vertex was used in addition to the hits in the TPC and ITS. This improves the track-momentum resolution and reduces the azimuthal dependence of the track reconstruction efficiency due to the non-uniform SPD response.
Accepted tracks were required to be measured with $0.15<\pT<100$~\GeVc\ in $|\eta|<0.9$, 
and to have at least 70 TPC space-points and no less than 80\% of the geometrically findable space-points in the TPC.

The single-track tracking efficiency was estimated from the detector response of HIJING~\cite{hijing} events reconstructed to detector level using 
GEANT3~\cite{geant3ref2} for the particle transport.
In the $0$--$10$\% centrality class, it is about 56\% at $0.15$~\GeVc, about $83$\% at $1.5$~\GeVc\ 
and then decreases to $81$\% at 3~\GeVc, after which it increases and levels off to about $83$\% at above 6.5~\GeVc.
For the $10$--$30$\% most central collisions, the tracking efficiency follows a similar $\pt$-dependence pattern, 
with absolute values of the efficiency that are $1$ to $2$\% higher compared to the $0$--$10$\% most central collisions.
The momentum resolution, which was estimated on a track-by-track basis using the covariance matrix of the track fit, is about $1\%$ at $1$~\GeVc\ 
and about $3$\% at $50$~\GeVc.
The contamination by secondary particles~\cite{ALICE-PUBLIC-2017-005} produced in particle-material interactions, conversions, and weak-decay products of long-lived particles is on the level of few percent.

\section{Correlation analysis}
\label{sec:correlationanalysis}

The two-dimensional associated per-trigger yield $Y(\Delta \varphi, \Delta \eta)$ measures the distribution of particles relative to the jet axes in bins of $\Delta \varphi$, $\Delta \eta$, event centrality, and trigger and associate transverse momenta $p_\mathrm{T,\,assoc}$~\cite{ABELEV:2013wsa}.
This distribution serves as the basis of the analysis and is formed using so-called same and mixed event correlations.
Correlations from the \textit{same event} are the actual correlations of trigger jets and associated particles, calculated for each selected event.
In the mixed event technique, jets are correlated with particles from a pool containing
tracks from different events with similar trigger jet $\pT$, vertex $z$, and centralities.
For vertex $z$, there are six bins in this pool, whose boundaries are given by $(-10, -5, -2, 0, 2, 5, 10)$ in cm.
The boundaries for the centrality percentile binning are given by $(0,1,2,3,4,5,10,20,30,40,50)$.

The mixed-event-corrected associated per-trigger yield for given jet \pT-range, associate \pT-range, and centrality selection is defined as
\eq{eq:baseCF}{Y(\Delta \varphi, \Delta \eta) = \frac{1}{N_\mathrm{trig}} \frac{\mathrm{d}^2 N_\mathrm{assoc}}{\mathrm{d} \Delta \eta \mathrm{d} \Delta \varphi}
= \frac{1}{N_\mathrm{trig}} \sum_{\mathrm{cent}, z}{\left(\frac{\mathrm{d}^2 N_\mathrm{same}}{\mathrm{d} \Delta \eta \mathrm{d} \Delta \varphi} \left/
  \alpha \frac{\mathrm{d}^2 N_\mathrm{mixed}}{\mathrm{d} \Delta \eta \mathrm{d} \Delta \varphi}\right.\right)},}
where the ratios in the sum are formed differentially in bins of centrality and vertex $z$. 

The factor $\alpha$ in Eq.~\ref{eq:baseCF} is chosen such that the mixed-event correlations are normalized to unity
in the region $|\Delta\eta| < 0.2$, $|\Delta\varphi| < 0.2$ around the near-side jet peak where the efficiency for pairs of parallel jets and associates is largest.
The contribution of the statistical uncertainty of this normalization to the total statistical uncertainty is negligible.
The finite tracking efficiency and the contamination by secondaries~(see Sec.~\ref{sec:setup}) are taken into account and a correction has been performed for associated tracks differentially in $\eta$, $\pt$, centrality, and vertex $z$ for same and mixed event correlations in Eq.~\ref{eq:baseCF}.
The efficiency maps were created using Monte Carlo simulations for the same track definition and detector conditions.
However, this correction turns out to be negligible for all observables except for the absolute jet-associated yields, because its effect mostly cancels in the used relative observables, which will be defined in Sec.~\ref{sec:observables}.

In addition to the correction for detector inhomogenities and acceptance effects, the correlation also needs to be corrected for background.
The underlying background for the chosen observables mainly consists of the uncorrelated particle background baseline from soft processes and the flow modulation in the correlation function.
The background was found to be independent of $\Delta\eta$ within $|\eta| < 0.9$~\cite{Adam:2016ows} and is therefore estimated as a function of $\Delta\varphi$ for the whole $\Delta\eta$-range as $B(\Delta \varphi)$. To avoid including parts of the jet signal, $B(\Delta \varphi)$ is calculated in $1.0 < |\Delta\eta| < 1.4$, where the contribution from the jet is expected to be small, based on measurements in pp collisions.

The background is directly subtracted from the correlation function.
The background-corrected per-trigger yield serves as a basis for all subsequent measurements. It is defined as
\eq{eq:Ycorr}{Y_\mathrm{corr}(\Delta \varphi, \Delta \eta) = Y(\Delta \varphi, \Delta \eta) - B(\Delta \varphi).}

 \begin{figure}[tbh!]
  \begin{center}
   \includegraphics[width=0.95\textwidth]{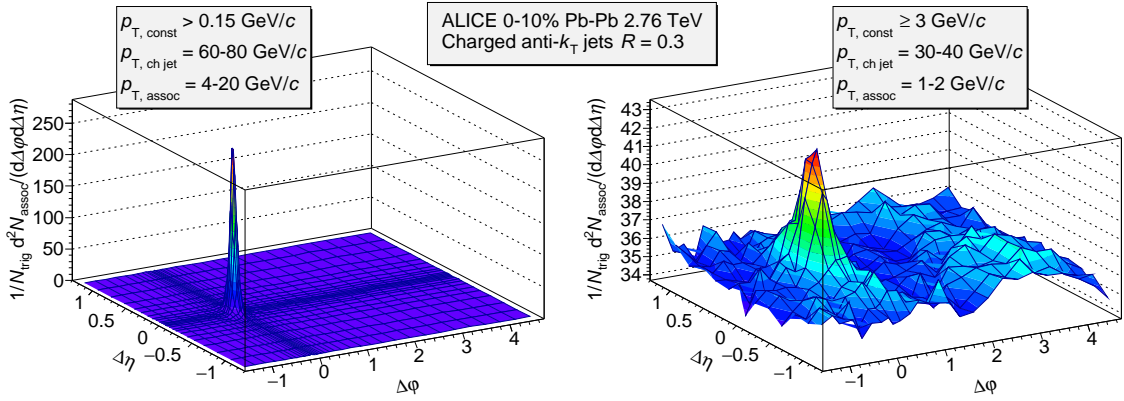}
   \caption{Illustration of per-trigger yields for the two different jet definitions (further discussed below): high-$\pT$ associates of jets with $p_\mathrm{T,\,const} \geq 0.15$ \GeVc~ and $p_\mathrm{T,\,ch\,jet} = 60$--$80$ \GeVc~ (left) and low-$\pT$ associates of jets with $p_\mathrm{T,\,const} \geq 3.0$ \GeVc~ and $p_\mathrm{T,\,ch\,jet} = 30$--$40$~\GeVc~ (right).
   No background subtraction was applied.}
   \label{fig:CF_Illustration}
  \end{center}
 \end{figure}

To illustrate the impact of the background on the per-trigger yields, the uncorrected per-trigger yields can be found in Fig.~\ref{fig:CF_Illustration} for high- and low-$\pT$ associates.
The background is nearly negligible for high-$\pT$ associates and it is sizeable for low-$\pT$ associates.
In the illustrated example for low-$\pT$ associates, the signal to signal+background ratio, i.e.\ the percentage of the signal in the measured observable,
is roughly 0.1 within a radius of $r<0.3$ around the near-side jet peak.
Note also that the background correction removes all $\Delta\eta$-independent correlations, including the away-side ridge
which is not investigated in the presented analysis.

\section{Jet reconstruction}
\label{sec:reconstruction}

The measurement of jets in heavy-ion collisions is challenging since a single event can contain multiple, possibly overlapping, jets
from independent hard nucleon--nucleon scatterings. In addition, low transverse momentum particles originating from soft processes lead to a fluctuating background
which strongly influences the jet reconstruction. The relative effect is largest for low-$\pT$ jets and most central events.
Consequently, jet reconstruction in heavy-ion collisions requires a robust jet definition,
and a procedure to correct for the presence of the large background~\cite{Cacciari:2010te}.

Jets were reconstructed using the anti-\kT\ or the \kT~algorithms~\cite{Cacciari:2008gp} in the \FastJet\ package~\cite{Cacciari:2011ma} with a resolution parameter of $R=0.3$.
Only those jets whose axis was reconstructed within $|\eta|<0.6$ were kept in the analysis to assure the nominal jet cone is fully contained within
the track acceptance of $|\eta|<0.9$. This limits the effect of the acceptance boundaries on the measured jet spectrum.
Jets reconstructed by the anti-\kT~algorithm were used to quantify signal jets,
while jets reconstructed by the \kT~algorithm were used to quantify the contribution from the underlying event~\cite{Abelev:2012ej}.

Two different jet definitions are used in this analysis: for measurements at high associate-$\pT$, jets are measured with a constituent cut $p_\mathrm{T,const} \geq 0.15$~\GeVc, measurements at low associate-$\pT$ are performed for jets measured with a constituent cut $p_\mathrm{T,const} \geq 3.0$~\GeVc.
Jets with $p_\mathrm{T,const} \geq 0.15$~\GeVc~are reconstructed using all charged particles available for jet reconstruction and, thus, the fragmentation bias is as small as possible.
This bias is caused by only including certain particles of the jet and could lead to a sample of harder fragmenting jets when leaving out particles at low $\pT$.
On the other hand, using all charged particles available for jet reconstruction also includes particles in the correlation analysis which were already used in the jet finding process.
The jet finding algorithm selects regions in momentum space with large energy flow. This implies that the distribution of charged particles inside the jet is biased. For example, the radial distribution of particles with respect to the jet axis will show a small depletion at distances just outside the jet cone radius $R$.
This particularly affects the shape of the jet, i.e.\ how the constituents are distributed relative to the jet axis, leading to an autocorrelation bias.

Therefore, the jets themselves and in particular their shapes are intimately connected to the jet definition.
For high-$\pt$ associates, the autocorrelation bias cannot be avoided and has to be accepted as a part of the jet definition.

Low-$\pT$ associates are broadly distributed up to large distances relative to the jet. Since the jet finding algorithm clusters the jets roughly
into cones with a nominal radius of $R=0.3$, it strongly affects the shape of the jet.
When measuring properties of low-$\pT$ associates, we avoid the autocorrelation bias by adapting the jet definition:
Trigger jets and associates can be decoupled by using jets with constituents \textit{above} a certain threshold and associates \textit{below} the threshold.
Therefore, for measurements at low associate-$\pT$, jets are reconstructed with $p_\mathrm{T,const} \geq 3$~\GeVc.
Using a geometrical matching procedure that is performed on two collections of the differently defined jets which are reconstructed in each event it was checked that the jet axes for both jet definitions do not strongly change.
For instance, for jets with $p_\mathrm{T,const} \geq 3$~\GeVc\ and $p_\mathrm{T,\,ch\,jet} > 30$~\GeVc\, the mean and standard deviation of the matched jet distance distribution are approximately given by $0.016$ and $0.014$, respectively.
However, it must be emphasized that these jet definitions select two different jet samples and that the autocorrelation bias was avoided here at the expense of a possible fragmentation bias.

The transverse momentum of reconstructed jets including constituents as low as $0.15$~\GeVc\ is affected by the contribution from the underlying event.  
In order to suppress the contribution of such background to the measurement of the jet momentum, we followed the approach 
described in \Refs{Abelev:2013kqa,Abelev:2012ej}, which addresses the average additive contribution to the jet momentum on a jet-by-jet basis.
The underlying background momentum density $\rho$ was estimated event-by-event using the median of $\jetraw/\Ajet$, 
where \jetraw\ is the uncorrected jet transverse momentum and $\Ajet$ is the area of jets reconstructed with the \kT~algorithm.

The average raw background momentum density $\left<\rho\right>$ decreases towards more peripheral collisions.
It is $\left<\rho\right>\approx110,\;65$, and $\;25$ \GeVc\ in the $0$--$10$\%, $10$--$30$\%, and $30$--$50$\% most central \PbPb\ collisions, respectively.
The background momentum density is a detector-level quantity that depends on the tracking efficiency and track definition.
For signal jets reconstructed with the anti-\kT\ algorithm and constituents above $0.15$~\GeVc, the background density scaled by the area of the reconstructed 
signal jet was subtracted from the raw reconstructed transverse momentum ($\jetraw$) of the signal jet according to $p_\mathrm{T,\,ch\,jet}=\jetraw-\rho\cdot\Ajet$.

Due to region-to-region variations of the background, the background-corrected jet transverse momenta are affected by residual fluctuations.
To give an estimate for these fluctuations for the jet definition used, cones with radius $R=0.3$ are randomly placed in each event.
In these cones, the track momenta are summed and the background is subtracted to calculate $\delta p_\mathrm{T}$:
\eq{eq:DeltaPt}{\delta\pt = \sum_\mathrm{cone}{p_\mathrm{T,\,track}} - \rho \cdot A,}
where $A$ is the area of the cone.

For the $0$--$10$\%, $10$--$30$\%, and $30$--$50$\% most central collisions, the standard deviation of the $\delta\pt$-distribution
as a measure for the magnitude of the fluctuations has been evaluated to $6.7$, $5.1$, and $3.3$~\GeVc, respectively.
Since the $\delta\pt$-distribution also contains the jet signal, the standard deviation of the full distribution is impacted by it.
A lower limit of these fluctuations is given by performing a Gaussian fit of the left-hand side of the $\delta\pt$-distribution.
The Gaussian widths were evaluated to $5.5$, $4.0$, and $2.3$~\GeVc\ for the $0$--$10$\%, $10$--$30$\%, and $30$--$50$\% most central collisions.
The sample of jets that only uses constituents above $3$~\GeVc\ is not corrected for the underlying event
as the constituent cut already strongly suppresses the contribution from the background such that it is negligible.

In addition to background fluctuations, also the finite detector resolution and single particle efficiency influence the measurement.
To quantify both effects, the ratio of reconstructed jet momentum $p_\mathrm{T,\,rec}$ and true jet momentum $p_\mathrm{T,\,true}$ was calculated
taking into account the detector resolution by using a response matrix and background fluctuations given by the $\delta\pT$ distributions.
The response matrix was created from Monte Carlo simulations for which the true jet momentum is known by geometrically matching particle-level PYTHIA jets with the corresponding detector-level jets reconstructed using a full detector model in GEANT3.
More detailed studies have been performed for jets on the same dataset in \Ref{Abelev:2013kqa}.

There are two effects contributing to the jet momentum resolution: detector effects and underlying event fluctuations.
The detector effects lead to a jet momentum response that is peaked at $p_\mathrm{T,\,rec} = p_\mathrm{T,\,true}$, but has a tail to lower values of detector level momentum due to tracking inefficiency.
The tracking efficiency changes by only a few percent from peripheral to central events.
Background fluctuations produce an approximately Gaussian response, with a width that depends strongly on centrality. The combined effect leads to a standard deviation in the jet momentum resolution of 30\% (20\%) for jets with $p_\mathrm{T,\,ch\,jet} = 30$~\GeVc\ and 27\% (27\%) for jets with $p_\mathrm{T,\,ch\,jet} = 120$~\GeVc\ for the $0$--$10$\% ($10$--$30$ and $30$--$50$\%) most central events.

It should be emphasized that $p_\mathrm{T,\,ch\,jet}$ refers to the jet transverse momentum at detector level, corrected for background only. Since within-event fluctuations
of the background are not corrected for, the mean of the given $p_\mathrm{T,\,ch\,jet}$-range is slightly higher than that of the underlying \textit{true} $\pT$ distribution for more central collisions where fluctations are dominant.
Hence, due to the steeply-falling jet spectrum, fluctuations lead to a shift of the spectrum to larger values.
For more peripheral collisions where detector effects are dominant, there is the opposite effect, i.e.\ the spectrum is shifted to smaller values.
The fraction of purely combinatorial jets in the momentum ranges used in the analysis was found to be negligible.

To give a rough estimate of the true jet populations for a given reconstructed jet momentum range, projections of the response matrices, introduced above, are used~\cite{ALICE-PUBLIC-2019-002}.
For measured $p_\mathrm{T,\,ch\,jet}$-distributions, approximate ranges are given in Tab.~\ref{tab:jetsamples} as a measure for the true jet momentum distributions.
The true populations are defined as the smallest possible ranges around the $p_\mathrm{T,\,ch\,jet}$-range in which at least 68\% of the jet population can be found.

\begin{table}
\caption{True jet populations $p_\mathrm{T,\,true}$ in \GeVc\ corresponding to given $p_\mathrm{T,\,ch\,jet}$-ranges for different event centrality classes. The ranges are given such that they contain at least 68\% of the jet population. The most probable values of the distributions are given in brackets.}
\centering
\begin{tabular}{l|lll|ll}
$p_\mathrm{T,\,const}$-cut & \multicolumn{3}{c}{0.15~\GeVc}  &  \multicolumn{2}{c}{3~\GeVc} \\
$p_\mathrm{T,\,ch\,jet}$ (GeV/$c$) & 40--60 & 60--80 & 80--120 & 30--40 & 40--60 \\ \hline
0--5\%     & 11--87 (44) & 22--111 (64) & 49--144 (94) & 7--59  (32) & 21--88 (46)  \\
5--10\%    & 11--86 (46) & 24--112 (66) & 52--146 (94) & 8--61  (32) & 22--89 (46)  \\
0--10\%    & 11--86 (46) & 25--113 (68) & 54--147 (94) & 10--63 (32) & 24--91 (48)  \\
10--30\%   & 13--86 (50) & 33--117 (70) & 63--149 (98) & 15--69 (32) & 30--94 (48)  \\
30--50\%   & 25--91 (52) & 47--118 (82) & 75--147 (98) & 23--73 (32) & 36--95 (52)  \\
\end{tabular}
\label{tab:jetsamples}
\end{table}

\section{Construction of PYTHIA baseline}
\label{sec:embeddedBaseline}

In this analysis, reconstructed detector-level PYTHIA-jets serve as vacuum baseline, because the size of the pp dataset at $\sqrt{s} = 2.76$~TeV is insufficient for this purpose.

To account for the fluctuations of the underlying event in \PbPb\ collisions, PYTHIA jets embedded in real \PbPb\ collisions are used as a reference.
Jets reconstructed in this reference dataset still show the same baseline jet properties but also include the effect of background fluctuations from the \PbPb\ event.
To create this reference dataset, the following procedure is applied.
Events are simulated with PYTHIA6 (Perugia-0~\cite{Skands:2010ak}, version 6.421) followed by transport in the detector using GEANT3 and full response simulation and reconstruction simulating the same detector conditions as in the \PbPb\ dataset.
The reconstructed tracks are embedded into \PbPb\ events, i.e.\ they are combined with tracks from \PbPb\ events.
In order to simulate the same conditions as in \PbPb, the tracking efficiency in pp is decreased to the level expected in \PbPb.
Since the tracking efficiency in pp is higher than in \PbPb, $2$\% of the PYTHIA tracks are randomly discarded before they are embedded~\cite{Abelev:2014ffa}.
Jet finding algorithms are applied to the PYTHIA event and also to the combined PYTHIA + \PbPb\ event.
Jets found in the combined event are only accepted for the reference dataset if they can be matched geometrically with those in the PYTHIA event.
A matched embedded jet needs to be less than $R=0.3$ away from a PYTHIA jet.


Due to the very high particle occupancy of the \PbPb\ collision system, the probability to reconstruct a PYTHIA jet in the embedded event
is much lower than the probability to reconstruct a jet of same momentum by overlapping a jet that already existed in the \PbPb\ event,
even after applying a geometrical matching procedure.
Therefore, without any further intervention, the embedded jet sample would consist mostly of \PbPb-jets overlapping low-$\pT$ PYTHIA jets.

Two approaches have been tested which ensure that the jet sample shows \PbPb-event-like fluctuations of a PYTHIA jet,
and not jets from the \PbPb\ event.
The analysis baseline technique uses a cut on the fraction of the jet \pT\ that originates from the matched jet in PYTHIA.
The applied cut values are motivated by the underlying true jet distribution that
shows two separated populations: jets mostly consisting of particles from PYTHIA or from \PbPb.
The cut value was chosen to achieve the best separation of the two distributions.
In the $0$--$10$\% most central collisions, it is required that at least 20\% of the jet constituents' \pT\ originate from the PYTHIA jet.
For more peripheral collisions, this fraction is increased to 25\%.
For jets with $p_\mathrm{T,\,const} \geq 3$~\GeVc, which were measured down to 30~\GeVc, a cut of 50\% is applied.
However, this procedure might impose a bias on the implicitly accepted background fluctuations.
Therefore, variations around these nominal values were considered for the evaluation of systematic uncertainties.
Alternatively, a jet veto technique has been used: an embedded jet is not accepted if it overlaps with an already existing jet
of sizeable transverse momentum $p_\mathrm{T,\,ch\,jet}$ in the \PbPb\ event. Several veto cut values between 15 and 40 GeV/$c$ were tested.
Eventually, it turns out that both approaches yield very similar results.
The reconstructed jets which survive the MC percentage cut serve as an input to the next analysis steps which are the same as in the data analysis.

\section{Observables}
\label{sec:observables}

In this analysis, two features of particle jets are probed in \PbPb\ collisions:
changes in the particle $\pT$ composition of jets and their radial distribution relative to the jet axis.

To probe relative changes in the charged particle $\pT$ composition of jets in a surrounding cone with $R=0.3$, the jet-associated yield ratio is measured.
The ratio is formed from the integrated jet-associated per-trigger yields $Y_\mathrm{PbPb}$ and $Y_\mathrm{emb}$ which represent the integrals of the per-trigger yield in the jet cone for a given $p_\mathrm{T,\,assoc}$-range as introduced in Eq.~\ref{eq:Ycorr}. Technically, the integral is the sum over the entries of all ($\Delta\eta$, $\Delta\varphi$)-bins whose center is within distances of up to $R=0.3$ around the jet axis in the background-corrected per-trigger yield histogram.

The jet-associated yield ratio is defined by $R_{Y} = Y_\mathrm{PbPb}/Y_\mathrm{emb}$.
It directly compares integrated jet-associated per-trigger yields in \PbPb\ to the same yields for embedded PYTHIA jets.
An enhancement or suppression in associated yields is directly seen as a deviation from unity in the ratio.


The relative radial particle distribution around the jet is directly derived from the jet-associated yields.
It shows the relative distribution of particle yields inside the jet cone.
Thus, it is a measure for the broadening or collimation of constituents with certain momenta in or around the jet cone.
As for the jet-associated yield ratio, this measurement is performed for high- and low-$\pt$ jet-associated yields.
The radial shape is normalized to represent a probability distribution.
It is defined in bins of $r = \sqrt{\Delta\eta^2 + \Delta\varphi^2}$, the distance to the jet axis, to exploit the radial symmetry of the jet peak.
In \Refs{Adam:2016a,Adam:2016b}, an asymmetric broadening of the near-side jet peak is observed in two-particle correlations. It is strongest for low associate and trigger momenta and vanishes for higher momenta.
Therefore, in the analysis presented here, the influence of this asymmetry on jet--hadron correlations was tested to check the radial symmetry of the jet peak.
Even for the lowest accessible jet and associated track momenta, no jet peak asymmetry was observed.
Measurements in $\Delta\eta$ and $\Delta\varphi$ lead to the same conclusions within statistical precision, which justifies the presentation of the jet radial shape in bins of $r$.
The correlation function which is used to obtain the shape is originally binned in $\eta$ and $\varphi$. The binning was chosen fine enough to avoid significant binning effects.

For a given centrality-bin, and trigger and associate $\pT$, it is defined by the following formula:
  \eq{def:YieldShape}{S(r_\mathrm{min}, r_\mathrm{max}) = \frac{1}{A} \int_{r_\mathrm{min}}^{r_\mathrm{max}} Y_\mathrm{corr} (r) \mathrm{d}r,}
where $Y_\mathrm{corr}(r)$ represents the background-corrected per-trigger yield, $r_\mathrm{min}$ and $r_\mathrm{max}$ the bin edges, and $A = \int_{0}^{r_\mathrm{range}} Y_\mathrm{corr} (r) \mathrm{d}r$ the integral for the self-normalization of the radial shape. The upper limit in the integral used for the self-normalization is chosen to reflect the different ranges of the shown radial shape and is $r_\mathrm{range} = 0.3$ for the jets with $p_\mathrm{T,\,const} \geq 0.15$~\GeVc\ and $r_\mathrm{range} = 0.9$ for jets with $p_\mathrm{T,\,const} \geq 3$~\GeVc.
The statistical uncertainty is calculated taking into account the self-normalization.

\section{Systematic uncertainties}
\label{sec:systematics}

\begin{table}[t]
\caption{Table of systematic uncertainties for jet-associated yields in \PbPb, embedded PYTHIA, and their ratio for high-$\pT$ associates (4--20~GeV/$c$) and low-$\pT$ associates (1--2~GeV/$c$) and  for the $0$--$10$\% most central collisions. Uncertainties are given as relative uncertainties in percentages.}
\centering
\begin{tabular}{l|l|lll|lll}
$p_\mathrm{T,\,assoc}$ (GeV/$c$) &  & \multicolumn{3}{c}{4--20}  &  \multicolumn{2}{c}{1--2} & \\
$p_\mathrm{T,\,ch\,jet}$ (GeV/$c$) &  Observable & 40--60 & 60--80 & 80--120 & 30--40 & 40--60 & \\ \hline
Background (\%)      & \PbPb                                 & 0.3--0.6      & 0.7--1.5      & 1.5--2.0 & 6.9      & 8.0       &  \\
                      & Embedded                               & 0.3--0.7      & 0.7--1.0      & 1.0--1.1 & 6.8      & 6.7       &  \\
                      & Ratio                                 & 0.4--0.7      & 0.1--0.7      & 0.4--1.6 & 6.9      & 9.6       &  \\ \hline
Mixed event correction (\%)  & \PbPb                         & 0.2      & 0.3      & 0.5 & 0.2      & 0.2       &  \\
                           & Embedded                            & 0.7      & 0.4      & 0.4 & 0.1      & $<$ 0.1       &  \\
                           & Ratio                            & 0.7      & 0.5      & 0.3 & 0.2      & 0.2       &  \\ \hline
Embedding (\%)                             & \PbPb                          & -      & -      & - & -      & -       &  \\
             & Embedded                          & 0.1--2.3      & 0.1--0.4      & 0.1--0.3 & 5.0      & 2.7       &  \\
                             & Ratio                          & 0.1--2.3      & 0.1--0.4      & 0.1--0.3 & 4.6      & 2.7       &  \\ \hline
Tracking efficiency (\%) & \PbPb                      & 4.0      & 4.0      & 4.0 & 4.0      & 4.0       &  \\
                               & Embedded                        & 4.0      & 4.0      & 4.0 & 4.0      & 4.0       &  \\
                                 & Ratio                      & -      & -      & - & -      & -       &  \\ \hline
Tracking PYTHIA (\%) & \PbPb                      & -      & -      & - & -      & -       &  \\
                               & Embedded                        & 2.0      & 2.0      & 2.0 & 2.0      & 2.0       &  \\
                                 & Ratio                      & 2.0      & 2.0      & 2.0 & 2.0      & 2.0       &  \\ \hline
PYTHIA vs.\ pp (\%)    & \PbPb                            & -          & -          & - & -          & -           & \\
        & Embedded                            & 5.0          & 5.0          & 5.0 & 2.0          & 2.0           & \\
        & Ratio                            & 5.0          & 5.0          & 5.0 & 2.0          & 2.0           & \\ \hline \hline
Total (\%)  &  \PbPb                    & 4.0--4.1      & 4.1--4.3      & 4.3--4.5 & 8.0      & 9.0       & \\ 
  &  Embedded                    & 6.8--7.2      & 6.8      & 6.8 & 9.8      & 8.7       & \\ 
  &  Ratio                    & 5.5--5.9      & 5.4--5.5      & 5.4--5.6 & 8.8      & 10.3       & \\
\end{tabular}
\label{tab:syst1}
\end{table}

\begin{table}[t]
\caption{Table of systematic uncertainties for jet radial shapes for high-$\pT$ associates (4--20~GeV/$c$) in \PbPb\ and embedded PYTHIA for the $0$--$10$\% most central collisions. Uncertainties are given as relative uncertainties in percentages.  Note that relative uncertainties grow for higher $r$ values.}
\centering
\begin{tabular}{l|lll|llll}
Data sample & \multicolumn{3}{c}{\PbPb}  &  \multicolumn{3}{c}{Embedded PYTHIA} & \\
$p_\mathrm{T,\,ch\,jet}$ (GeV/$c$) & 40--60 & 60--80 & 80--120 & 40--60 & 60--80 & 80--120 &  \\ \hline
Background (\%)                                      & 0.1--6.5      & 0.1--13.0      & 0.1--19.2& 0.0--6.9      & 0.0--10.8      & 0.0--14.5       &  \\
Mixed event corr. (\%)                          & $<$ 0.1      & $<$ 0.1      & $<$ 0.1& $<$ 0.1      & $<$ 0.1      & $<$ 0.1       &  \\
Embedding (\%)                                       & -      & -      & -& 1.0--13.9      & 0.4--3.1      & 0.1--0.8       &  \\
PYTHIA vs.\ pp (\%)                                   & -          & -          & -& 2.0          & 2.0          & 2.0           & \\ \hline
Total (\%)                       & 0.1--6.5      & 0.1--13.0      & 0.1--19.2& 2.2--15.7      & 2.0--11.5      & 2.0--14.6       & \\
\end{tabular}
\label{tab:syst2}
\end{table}

\begin{table}[t]
\caption{Table of systematic uncertainties for jet radial shapes for low-$\pT$ associates (1--2~GeV/$c$, 2--3~GeV/$c$) in \PbPb\ and embedded PYTHIA for jets with $p_\mathrm{T,\,ch\,jet}= $ 40--60 \GeVc\ and for the $0$--$10$\% most central collisions. Uncertainties are given as relative uncertainties in percentages. Note that relative uncertainties grow for higher $r$ values.}
\centering
\begin{tabular}{l|ll|lll}
Data sample & \multicolumn{2}{c}{\PbPb}  &  \multicolumn{2}{c}{Embedded PYTHIA} & \\
$p_\mathrm{T,\,assoc}$ (GeV/$c$) & 1--2 & 2--3 & 1--2 & 2--3 &  \\ \hline
Background (\%)                                       & 1.6--7.5      & 0.4--8.8& 2.2--11.9      & 1.0--4.2       &  \\
Mixed event corr. (\%)                           & $<$ 0.1      & $<$ 0.1& $<$ 0.1      & $<$ 0.1       &  \\
Embedding (\%)                                       & -      & -& 1.2--7.4      & 0.8--11.3       &  \\
PYTHIA vs.\ pp (\%)                                    & -      & -& 2.0--10.0      & 2.0--10.0       &  \\ \hline
Total (\%)                        & 1.6--7.5      & 0.4--8.8& 6.2--13.0      & 4.5--15.7       & \\
\end{tabular}
\label{tab:syst3}
\end{table}

Several sources of systematic uncertainties contribute to the full uncertainty of the measurement and the evaluated individual uncertainties are combined using a quadratic sum, assuming they are uncorrelated.
Uncertainties for the following analysis aspects have been taken into account: the non-jet-related background correction technique,
the mixed-event correction, the selection of embedded jets, the tracking efficiency, and the impact of using a PYTHIA reference instead of a measured reference in pp at the same energy.
The uncertainties are partly correlated point-to-point.
The discussed uncertainties are summarized in Tabs.~\ref{tab:syst1}--\ref{tab:syst3}.

To correct for the non-jet-correlated background in the correlation function, the background is evaluated on the sidebands and subtracted in $\Delta\varphi$, as described in Sec.~\ref{sec:correlationanalysis}.
Different underlying background methods for the correlation functions have been tested:
for systematic uncertainties, the definition of the sideband range was varied to $1.1 < |\Delta\eta| < 1.3$ instead of $1.0 < |\Delta\eta| < 1.4$.
In addition, a simpler method that approximates the background by a constant baseline ($B(\Delta\varphi = \mathrm{const})$) has been used.

The mixed-event acceptance/inhomogenity correction is a small correction.
Two variations are considered for systematic uncertainties. First, the mixed-event correction is calculated inclusively for all $\Delta\varphi$.
Second, the normalization of the mixed-event correlations is performed for $|\Delta\eta| < 0.3$ and full $|\Delta\varphi|$ instead of using the plateau
in $|\Delta\eta| < 0.2$ and $|\Delta\varphi| < 0.2$.

In the embedding, a cut motivated by studying the underlying true jet distributions is applied on the fraction of jet \pT\ originating from the PYTHIA event,
as described in Sec.~\ref{sec:embeddedBaseline}.
Instead of cutting at 20\% for 0--10\% centrality, and 25\% for other centralities, the cut is varied to 15\% and 25\% for 0--10\% centrality,
and to 20\% and 30\% for other centralities.
As described above, for jets with $p_\mathrm{T,\,const} \geq 3$~\GeVc\ a baseline cut value of 50\% is used.
For systematic variation, the cut is performed at 15\% and 60\% for 0--10\% centrality, 20\% and 60\% for other centralities.

The detector has a finite single track reconstruction efficiency, which is only known with finite precision.
Since all observables are corrected for the tracking efficiency, they are all directly affected by its uncertainty.
Detailed studies of the tracking efficiency uncertainty have been performed to evaluate the size of its systematic uncertainty~\cite{Abelev:2014ffa, Abelev:2013kqa}.
The studies indicate that the (absolute) uncertainty is $4$\% for \PbPb\ collisions,
mainly due to an imperfect description of the ITS-TPC matching efficiency.
Another uncertainty from the tracking efficiency correction enters this analysis due to the usage of PYTHIA simulations.
The tracking efficiency of the PYTHIA data is artificially lowered by $2\%$ before embedding to account for the lower tracking efficiency in \PbPb\ collisions.
As a conservative estimate, a relative uncertainty of $100\%$ is assigned to this value.
Both components of the tracking efficiency uncertainty are taken into account as independent contributions to the uncertainty, i.e.\ added in quadrature to the full uncertainty.
These uncertainties are directly used as uncertainties for the yields, see Tab.~\ref{tab:syst1}. For the jet-associated yield ratio, the uncertainty on the tracking efficiency in \PbPb\ cancels, because it is correlated in \PbPb\ and the embedded PYTHIA reference.
For the radial shape distribution, a change in the tracking efficiency has no impact either, since these observables are relative quantities that do not depend on the global magnitude of the tracking efficiency.
As an alternative approach to estimate the impact of these two uncertainties of the tracking efficiencies on the observables, the full analysis was redone using corrections that assume the above given lower tracking efficiencies.
There was no significant impact on the presented results.

Finally, an uncertainty is assigned since PYTHIA is used as a baseline instead of a measured pp reference.
Including this uncertainty, the conclusions are also valid for a pp reference and not only for an embedded PYTHIA reference.
In order to do so, the presented observables were calculated and compared for PYTHIA events and pp collisions at 7 TeV. Within the statistical precision of this comparison, it is only possible to give an estimate for the inclusive $p_\mathrm{T,\,ch\,jet}$-range.
The relative deviations of each observable between both datasets enter directly as a systematic uncertainty and are on the level of a few percent, cf.\ Tabs.~\ref{tab:syst1}--\ref{tab:syst3}.

\section{Results}
\label{sec:results}

\begin{figure}[tbh!]
\begin{center}
  \includegraphics[width=0.49\textwidth]{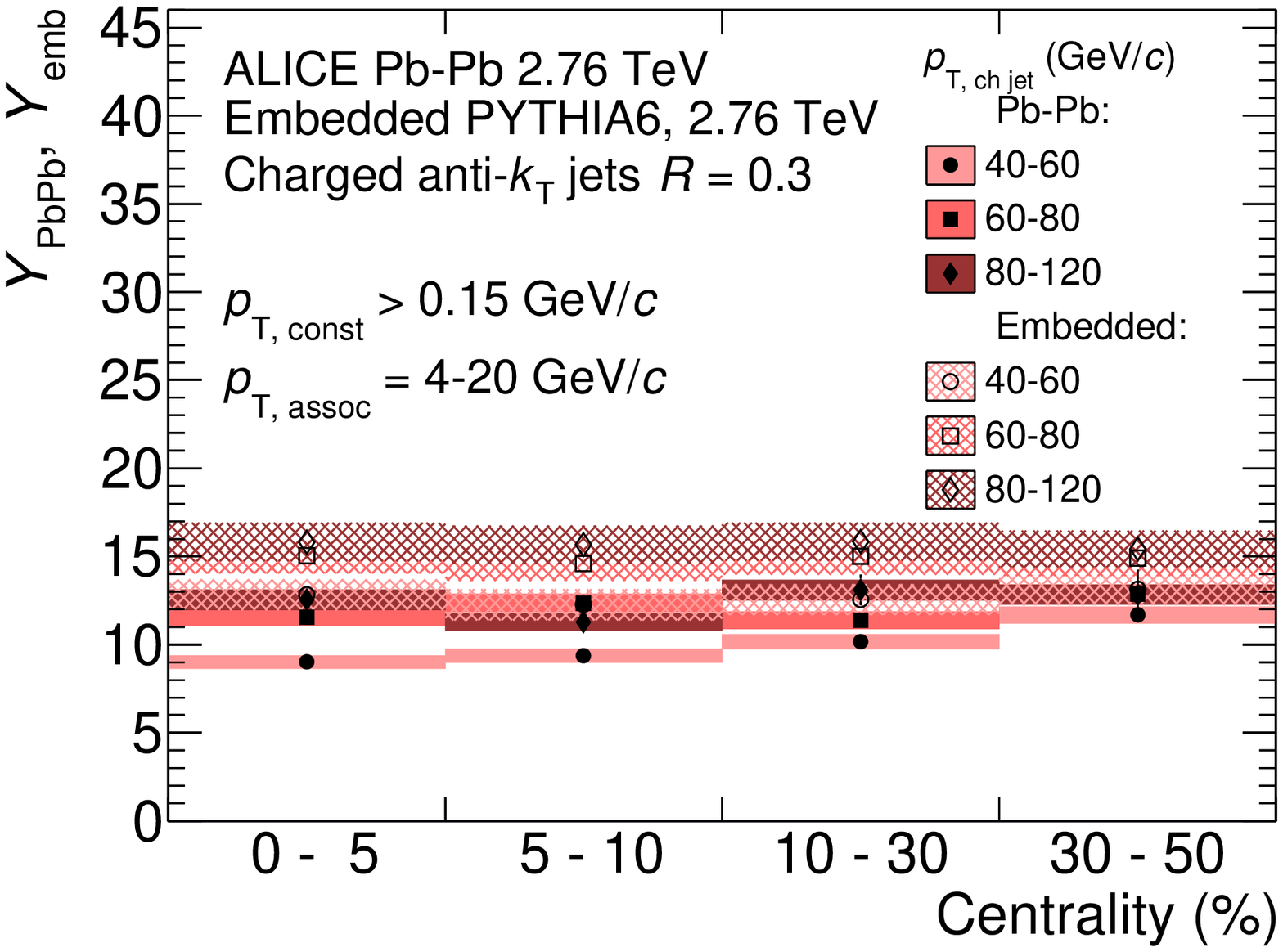}
  \includegraphics[width=0.49\textwidth]{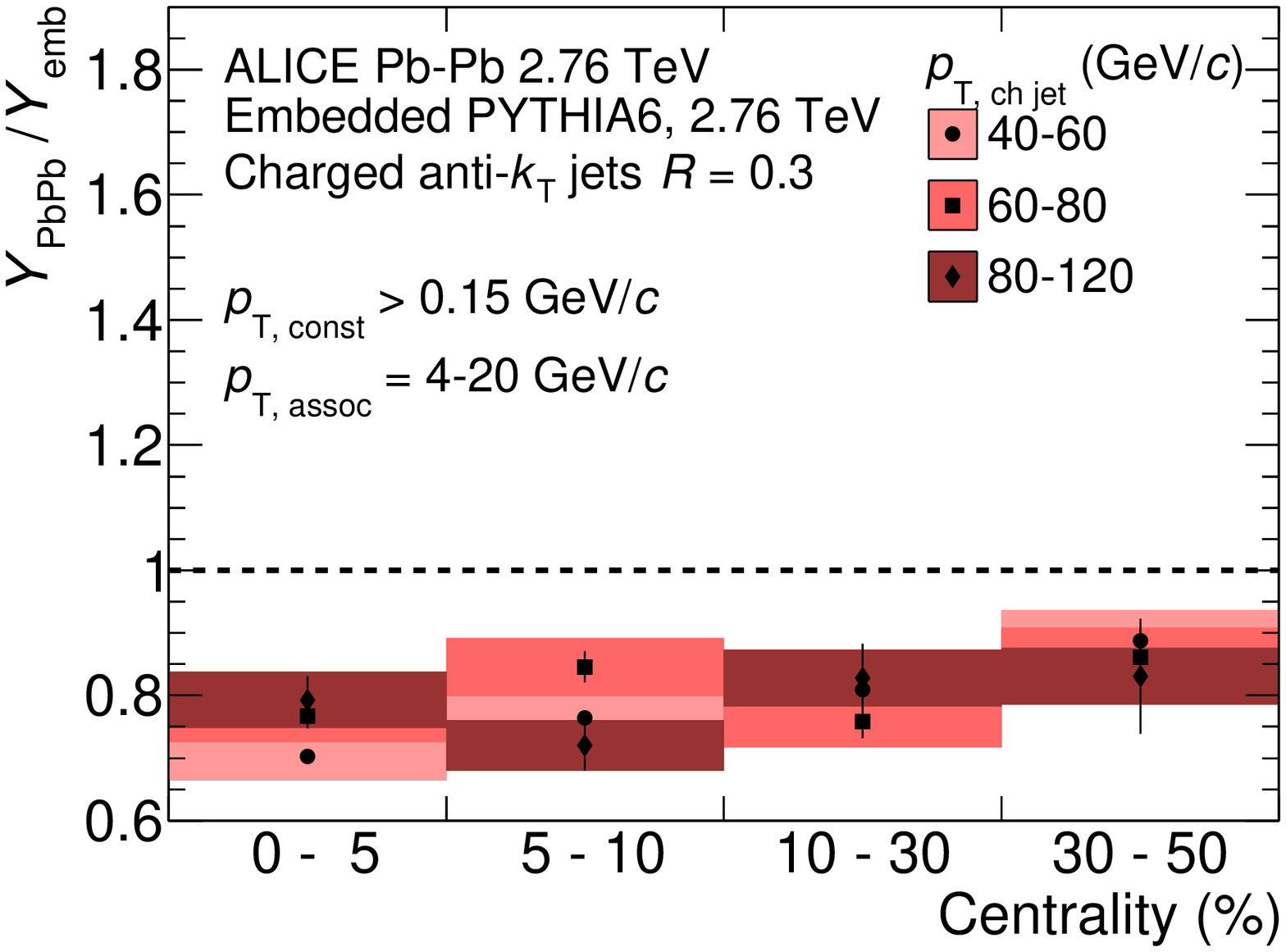}
  \caption{Centrality dependence of jet-associated yields (left) and yield ratios (right) for high-\pT\ associates. Boxes represent systematic uncertainties, error bars represent statistical uncertainties. Observables are corrected for acceptance and background effects.}
 \label{fig:1}
 \end{center}
\end{figure}

\begin{figure}[tbh!]
\begin{center}
  \includegraphics[width=0.49\textwidth]{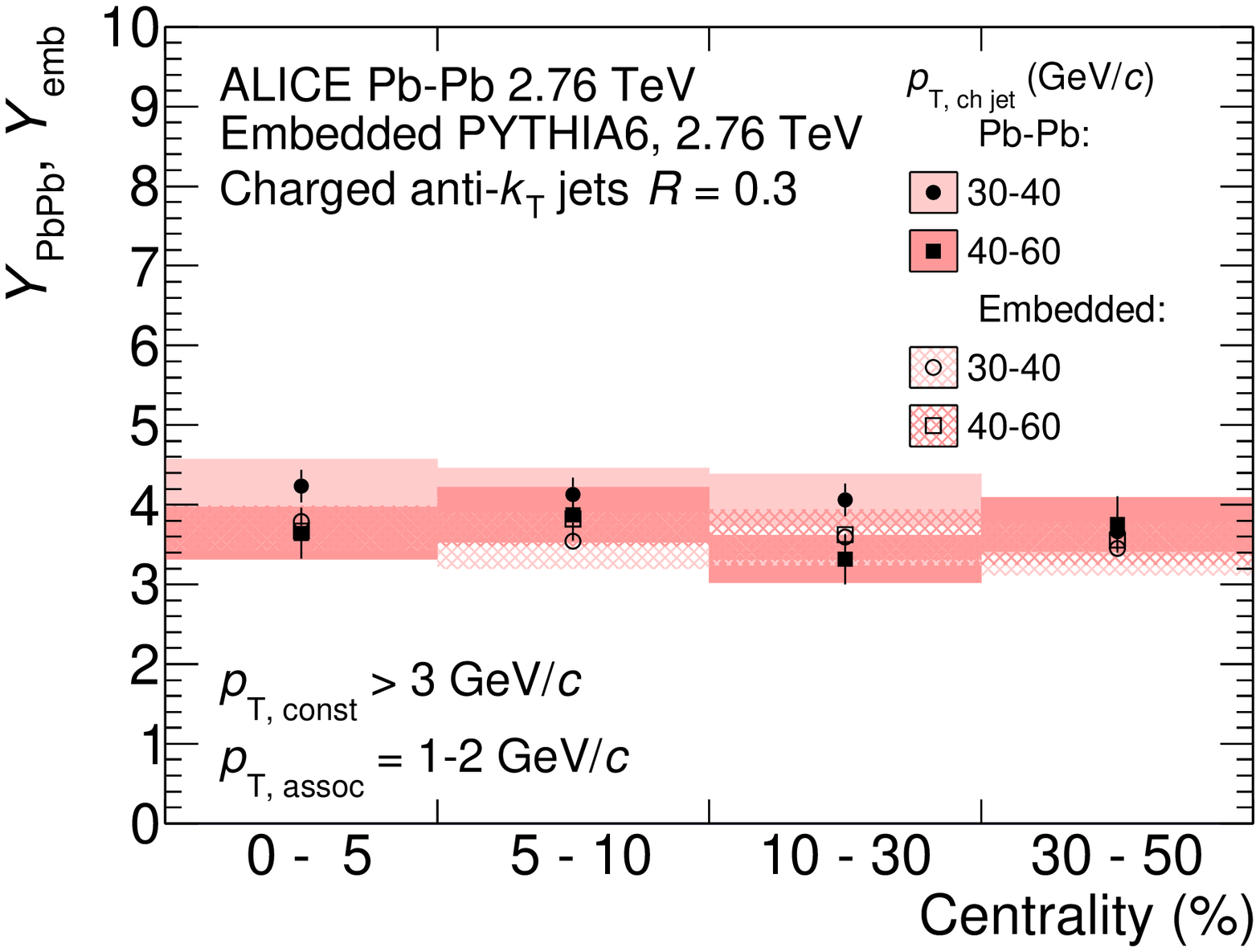}
  \includegraphics[width=0.49\textwidth]{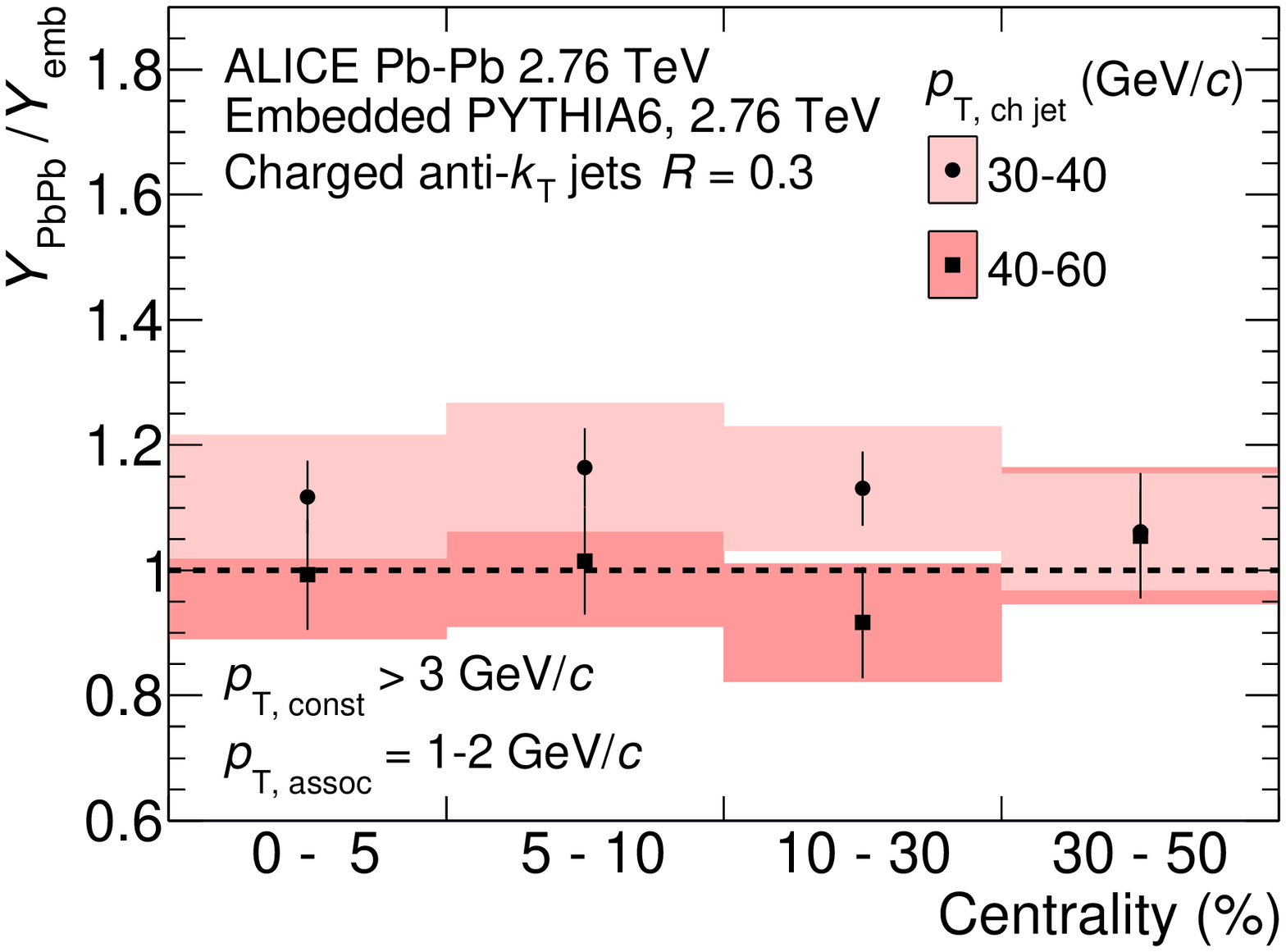}
  \caption{Centrality dependence of jet-associated yields (left) and yield ratios (right) for low-\pT\ associates. Boxes represent systematic uncertainties, error bars represent statistical uncertainties. Observables are corrected for acceptance and background effects.}
 \label{fig:2}
 \end{center}
\end{figure}


Figures \ref{fig:1} and \ref{fig:2} depict the jet-associated yields (left) and yield ratios (right) for high-\pT\ and low-\pT\ associated particles, respectively.
Both quantities are shown as a function of event centrality and for several selected jet transverse momenta.

The jet-associated yield ratio shows a suppression with a significance of several standard deviations in the centrality range 0--50\% for the considered high-$\pT$ associated particles.
In the probed jet momentum range, no significant $p_\mathrm{T,\,ch\,jet}$-dependence is observed.
The centrality-dependent linear slope of the distribution for $p_\mathrm{T,\,ch\,jet}= $ 40--60 \GeVc\ is more than one standard deviation away from zero, taking into account statistical and systematic uncertainties added in quadrature, indicating that there is a slightly stronger suppression for more central collisions in this case.
As a cross check, the same observable was also measured for jets with several higher minimum $p_\mathrm{T,\,const}$-cuts, i.e.\ 1, 2, and 3 \GeVc, which are less affected by the underlying event.
They lead to similar conclusions.

The jet-associated yield ratio for low-$\pT$ associates has much larger statistical and systematic uncertainties than the ratio of high-$\pT$ constituents,
thus it is not possible to draw a definite conclusion.

\begin{figure}[tbh!]
\begin{center}
  \includegraphics[width=0.49\textwidth]{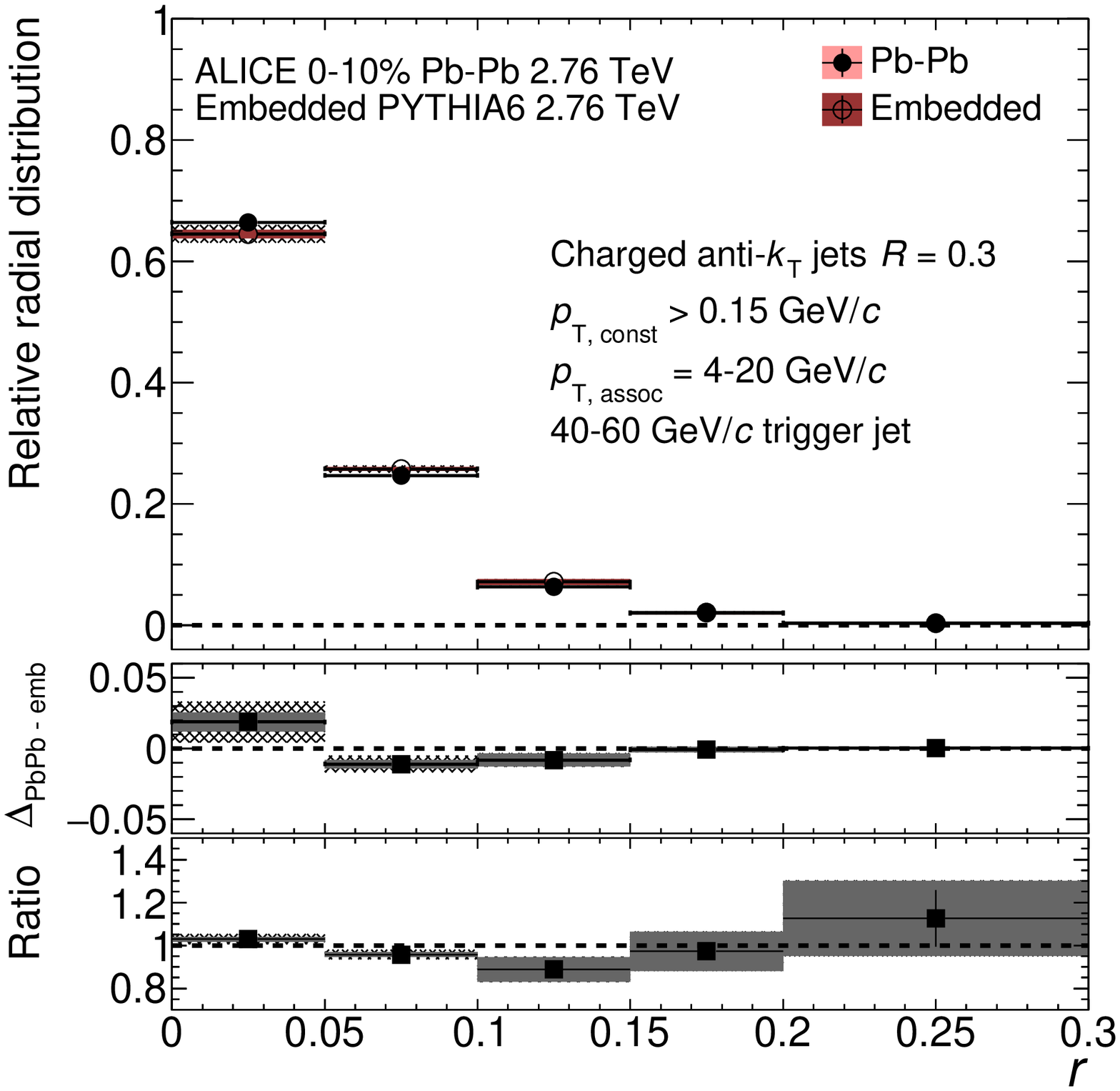}
  \includegraphics[width=0.49\textwidth]{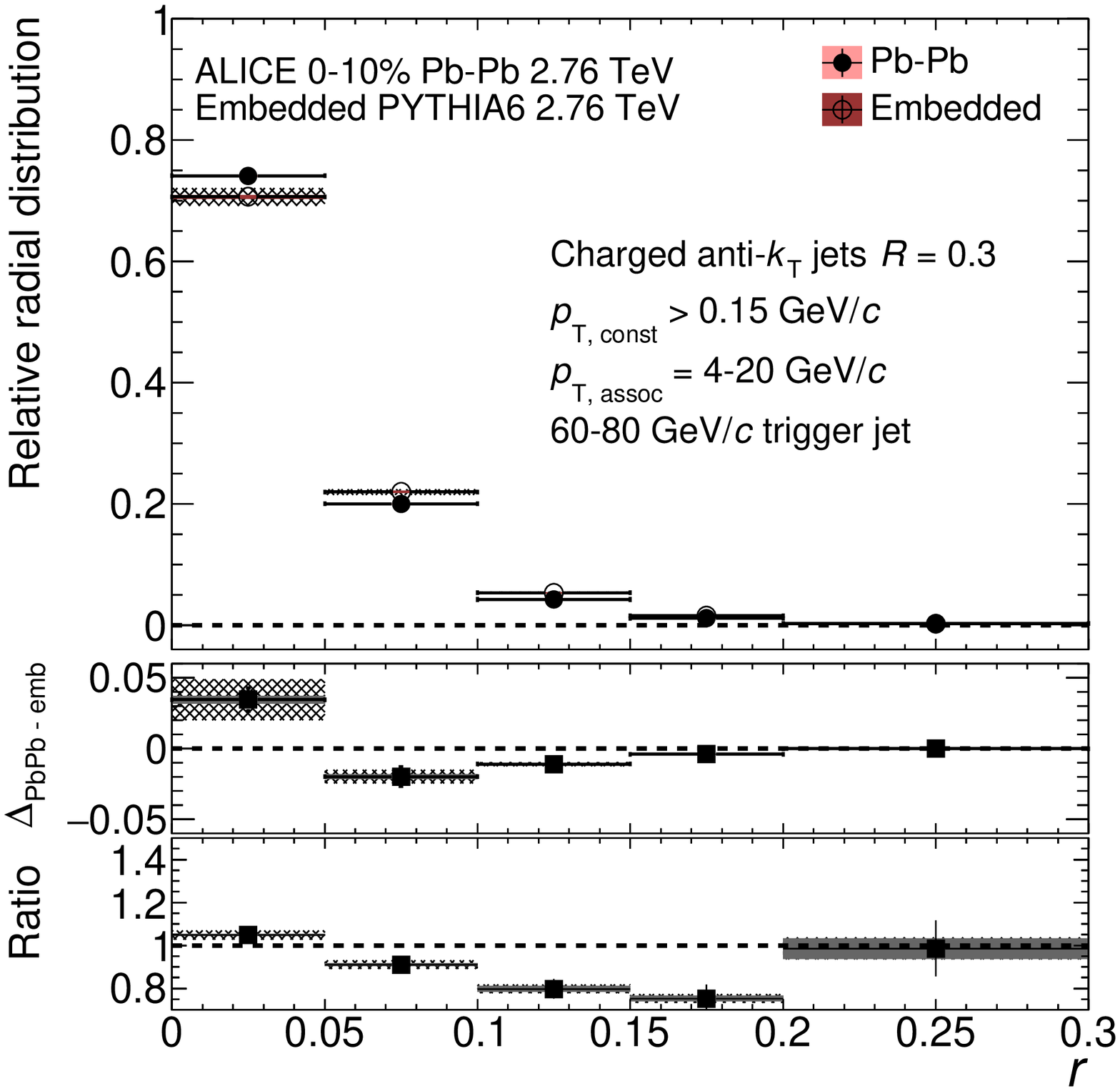}
  \includegraphics[width=0.49\textwidth]{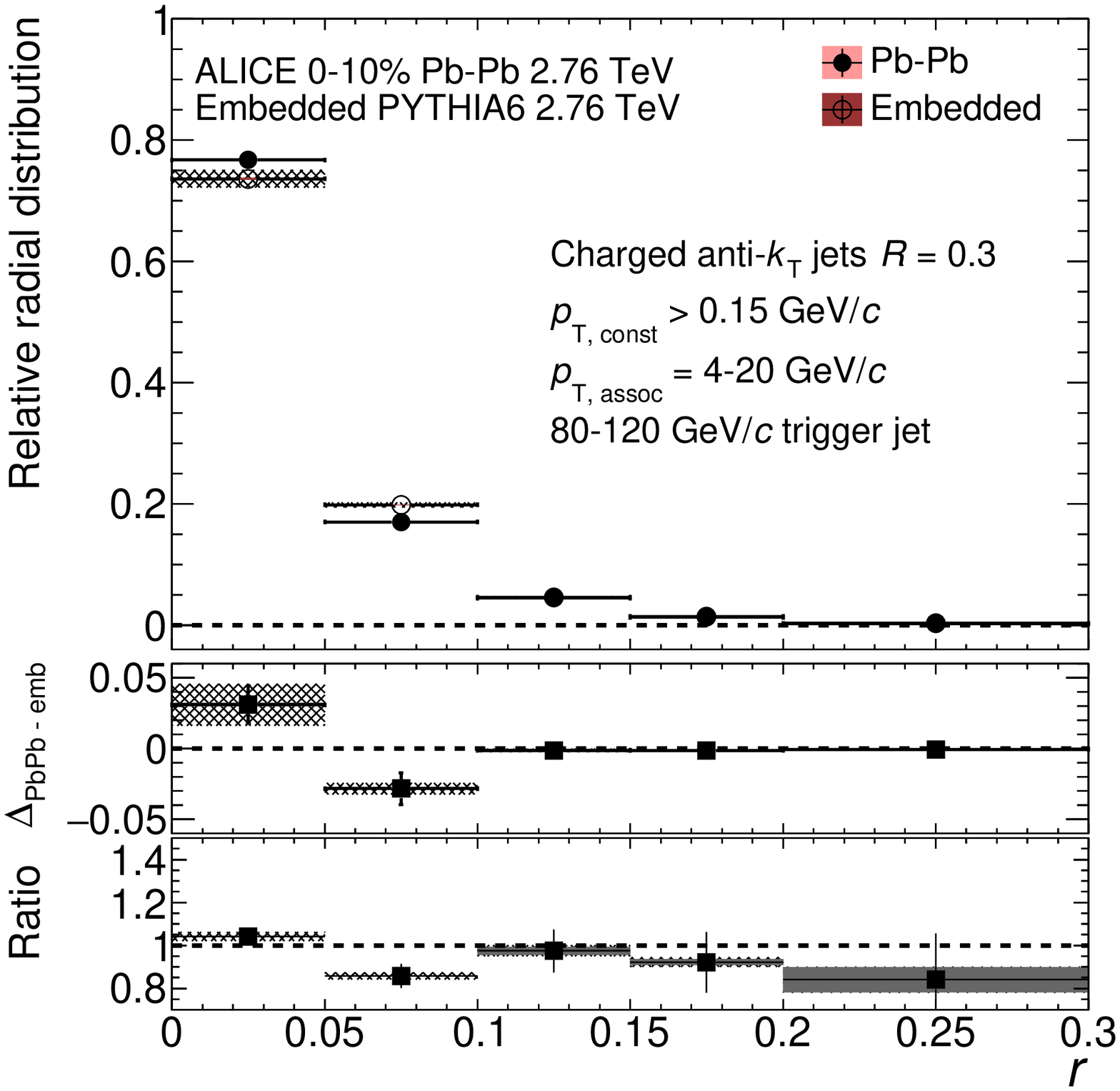}
  \caption{Jet relative radial shape distributions, differences, and ratios for the 0--10\% most central collisions for high-$\pT$ constituents, shown for different jet transverse momenta.
  Boxes represent systematic uncertainties, shaded boxes include uncertainties from PYTHIA/pp comparison, and error bars represent statistical uncertainties. Observables are corrected for acceptance and background effects.}
 \label{fig:3}
 \end{center}
\end{figure}

\begin{figure}[tbh!]
\begin{center}
  \includegraphics[width=0.49\textwidth]{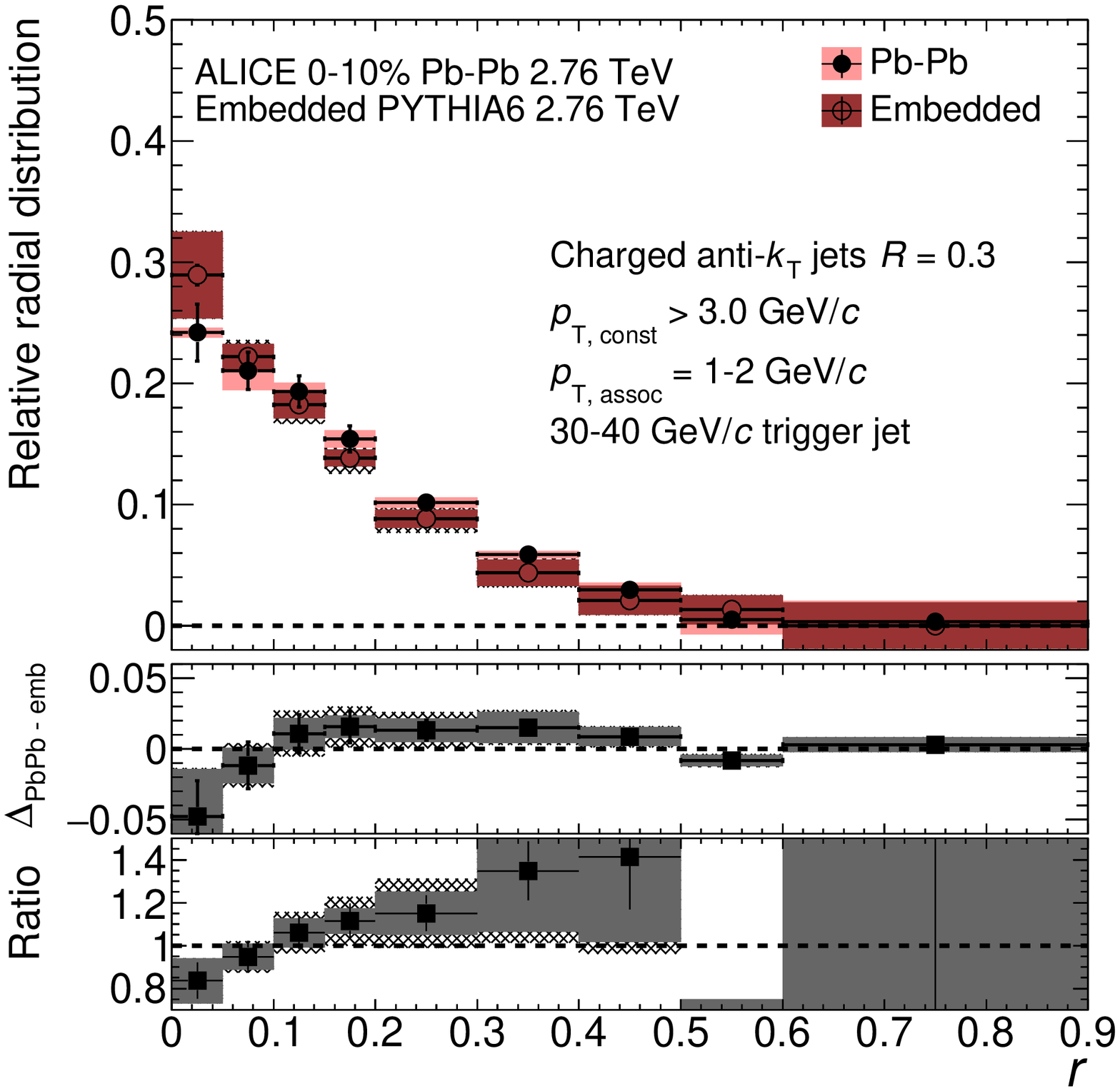}
  \includegraphics[width=0.49\textwidth]{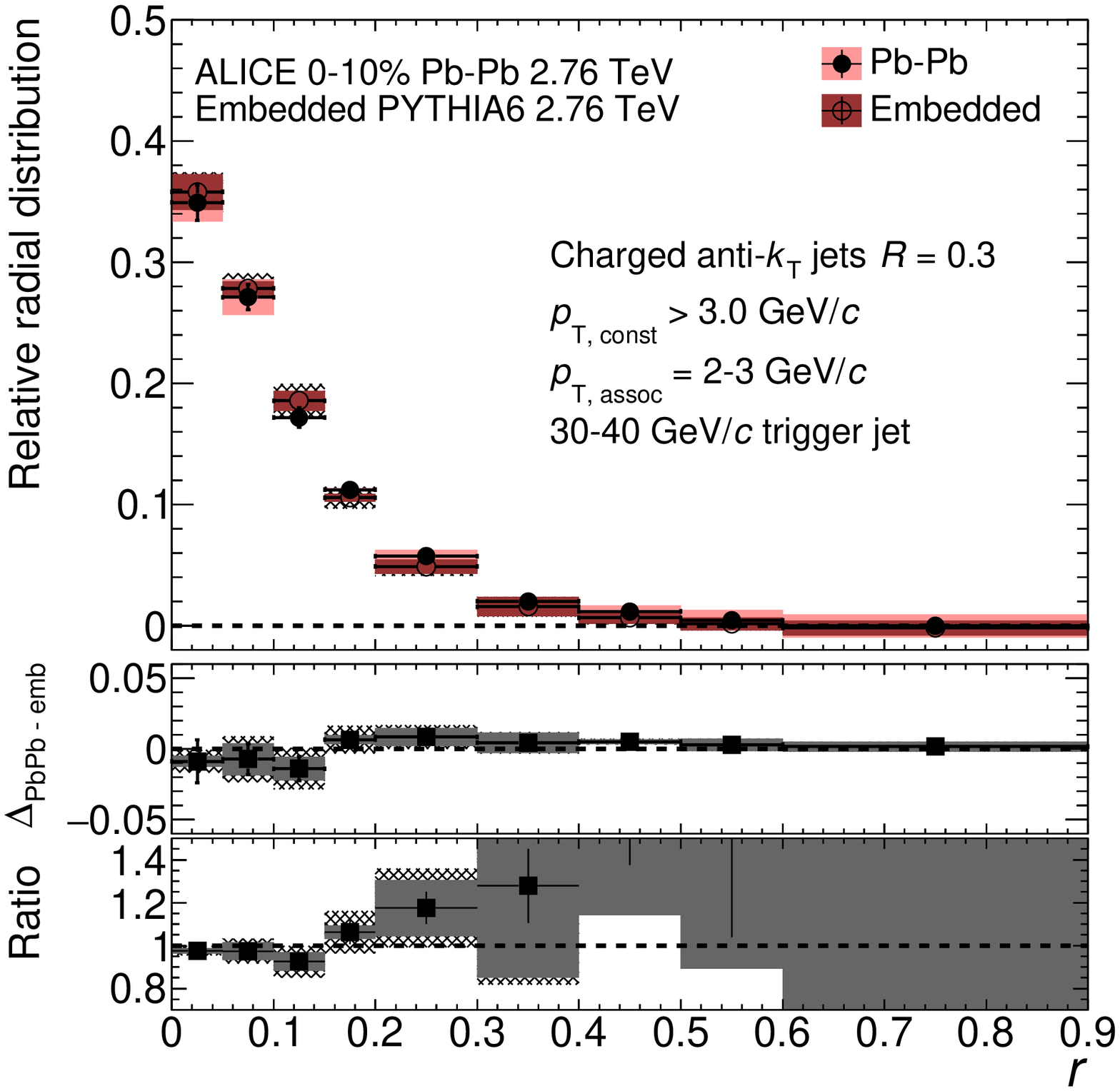}
  \caption{Jet relative radial shape distributions, differences, and ratios for the 0--10\% most central collisions for two different low-$\pT$ constituent ranges.
  Boxes represent systematic uncertainties, shaded boxes include uncertainties from PYTHIA/pp comparison, and error bars represent statistical uncertainties. Observables are corrected for acceptance and background effects.
  The $y$-axis scale of the ratio is chosen to focus on $r<0.3$, where the deviation of the ratio from unity is significant.
  }
 \label{fig:4}
 \end{center}
\end{figure}

The measured jet relative radial shapes are presented in Figs.~\ref{fig:3} and \ref{fig:4}. The top panels show the self-normalized distributions,
the difference and the ratio of the shapes in \PbPb\ and embedded PYTHIA can be found in the two lower panels.
The jet radial shapes of high-$\pT$ associates are measured for $p_\mathrm{T,\, ch\,jet}=$ 40--60~\GeVc, 60--80~\GeVc, and 80--120~\GeVc.
Shapes of low-$\pT$ associates are presented for jets with $p_\mathrm{T,\,ch\,jet} =$ 30--40~\GeVc\ and $p_\mathrm{T,\,const} > $3~\GeVc\ for associates
with $p_\mathrm{T,\,assoc} = $ 1--2~\GeVc\ and $p_\mathrm{T,\,assoc} = $ 2--3~\GeVc.

In general, the radial shape measurements indicate that all jet-associated yields are similarly distributed relative to the jet axis in \PbPb\ and embedded PYTHIA.
The yields of high-$\pT$ associates appear to be slightly more collimated near the core for jets in \PbPb, though the absolute effect is small.
While the shape is not significantly changed for jet transverse momenta between $40$ and $60$~\GeVc\ in \PbPb\ compared to the reference, there is a visible collimation for higher jet momenta above $60$~\GeVc. This can be seen best in the difference distributions $\Delta_{\mathrm{PbPb} - \mathrm{emb}}$ of Fig.~\ref{fig:3} which show that a larger fraction of the associated yield can be found near the core in \PbPb\ collisions.

The ratio distributions show that the collimation effect persists up to $r=0.2$, which is best visible for jets with $p_\mathrm{T,\,ch\,jet} =$ 60--80~\GeVc.
In the CMS measurement~\cite{CMS:2016a}, no significant change of the near-side jet peak width is observed in \PbPb\ for high-\pT\ associates and jets above 120~\GeVc.
However, the magnitude of the effect observed here is compatible with the observations within uncertainties.
Also note that the CMS data hints as well to a small collimation of the peak for higher-$\pT$ associates (4--8~\GeVc). 
Possible effects which might lead to a collimation include a relative change in the quark/gluon content in \PbPb\ compared to the reference~\cite{Spousta:2015fca}, as well as  a suppression of large-angle soft radiation in the coherent jet energy loss picture~\cite{CasalderreySolana:2010eh,CasalderreySolana:2012ef}.
Low-$\pT$ jet-associated yields presented in Fig.~\ref{fig:4} are measured up to a distance of $r=0.9$ relative to the jet
since in this case the associates are decoupled from the trigger jets.

For $p_\mathrm{T,\,assoc} = $ 1--2~\GeVc, a hint of a broadening of the radial shape is observed for jets with momenta between $30$ and $40$~\GeVc\ for the given definition. The broadening is visible in the difference distribution of the left plot in Fig.~\ref{fig:4}: in \PbPb\ collisions, a smaller fraction of particles can be found directly next to the jet axis.
For higher associate transverse momenta, i.e.\ $p_\mathrm{T,\,assoc} = $ 2--3~\GeVc, there is no significant modification of the low-$\pT$ radial shape of jets in \PbPb\ collisions within the large current experimental uncertainties.
A robust measurement of this observable for $p_\mathrm{T,\, ch\,jet}=$ 40--60~\GeVc\ or higher momenta is not possible due to the insufficient size of the dataset.
For higher jet momenta above 120~\GeVc, CMS measures a significant broadening of the near-side jet peak.

\section{Summary}
\label{sec:summary}
The presented results constitute the first attempt to study jet--hadron correlations with track-based jets down to transverse momenta of 30~\GeVc\ in \PbPb\ collisions --- a challenging regime due to the large underlying event and its fluctuations.
The jet radial shapes and the change in the particle $\pT$ composition were measured in \PbPb\ collisions at $\snn = 2.76$~TeV for high- and low-$\pT$ associates and compared to embedded PYTHIA simulations.
The number of high-$\pt$ associates in \PbPb\ collisions is suppressed compared to the reference by roughly 30 to 10\%, depending on centrality.
The radial particle distribution relative to the jet axis shows a moderate modification in \PbPb\ collisions with respect to PYTHIA.
High-$\pT$ associate particles are slightly more collimated in \PbPb\ collisions compared to the reference.
For jets with $p_\mathrm{T,\,const} \geq 3$~\GeVc, the radial distributions of low-$\pT$ associates were measured.
A hint of a broadening of the low-$\pT$ radial shapes is observed for $p_\mathrm{T,\,assoc} = $ 1--2~\GeVc. 
The shape for $p_\mathrm{T,\,assoc} = $ 2--3~\GeVc\ does not show a significant modification within its large uncertainties.
The results are in line with both previous jet--hadron-related measurements from the CMS Collaboration and jet shape measurements from the ALICE Collaboration at higher $\pt$
and add further support for the established picture of in-medium parton energy loss.

\ifpreprint
\iffull
\newenvironment{acknowledgement}{\relax}{\relax}
\begin{acknowledgement}
\section*{Acknowledgements}

The ALICE Collaboration would like to thank all its engineers and technicians for their invaluable contributions to the construction of the experiment and the CERN accelerator teams for the outstanding performance of the LHC complex.
The ALICE Collaboration gratefully acknowledges the resources and support provided by all Grid centres and the Worldwide LHC Computing Grid (WLCG) collaboration.
The ALICE Collaboration acknowledges the following funding agencies for their support in building and running the ALICE detector:
A. I. Alikhanyan National Science Laboratory (Yerevan Physics Institute) Foundation (ANSL), State Committee of Science and World Federation of Scientists (WFS), Armenia;
Austrian Academy of Sciences, Austrian Science Fund (FWF): [M 2467-N36] and Nationalstiftung f\"{u}r Forschung, Technologie und Entwicklung, Austria;
Ministry of Communications and High Technologies, National Nuclear Research Center, Azerbaijan;
Conselho Nacional de Desenvolvimento Cient\'{\i}fico e Tecnol\'{o}gico (CNPq), Universidade Federal do Rio Grande do Sul (UFRGS), Financiadora de Estudos e Projetos (Finep) and Funda\c{c}\~{a}o de Amparo \`{a} Pesquisa do Estado de S\~{a}o Paulo (FAPESP), Brazil;
Ministry of Science \& Technology of China (MSTC), National Natural Science Foundation of China (NSFC) and Ministry of Education of China (MOEC) , China;
Croatian Science Foundation and Ministry of Science and Education, Croatia;
Centro de Aplicaciones Tecnol\'{o}gicas y Desarrollo Nuclear (CEADEN), Cubaenerg\'{\i}a, Cuba;
Ministry of Education, Youth and Sports of the Czech Republic, Czech Republic;
The Danish Council for Independent Research | Natural Sciences, the Carlsberg Foundation and Danish National Research Foundation (DNRF), Denmark;
Helsinki Institute of Physics (HIP), Finland;
Commissariat \`{a} l'Energie Atomique (CEA), Institut National de Physique Nucl\'{e}aire et de Physique des Particules (IN2P3) and Centre National de la Recherche Scientifique (CNRS) and Rl\'{e}gion des  Pays de la Loire, France;
Bundesministerium f\"{u}r Bildung, Wissenschaft, Forschung und Technologie (BMBF) and GSI Helmholtzzentrum f\"{u}r Schwerionenforschung GmbH, Germany;
General Secretariat for Research and Technology, Ministry of Education, Research and Religions, Greece;
National Research, Development and Innovation Office, Hungary;
Department of Atomic Energy Government of India (DAE), Department of Science and Technology, Government of India (DST), University Grants Commission, Government of India (UGC) and Council of Scientific and Industrial Research (CSIR), India;
Indonesian Institute of Science, Indonesia;
Centro Fermi - Museo Storico della Fisica e Centro Studi e Ricerche Enrico Fermi and Istituto Nazionale di Fisica Nucleare (INFN), Italy;
Institute for Innovative Science and Technology , Nagasaki Institute of Applied Science (IIST), Japan Society for the Promotion of Science (JSPS) KAKENHI and Japanese Ministry of Education, Culture, Sports, Science and Technology (MEXT), Japan;
Consejo Nacional de Ciencia (CONACYT) y Tecnolog\'{i}a, through Fondo de Cooperaci\'{o}n Internacional en Ciencia y Tecnolog\'{i}a (FONCICYT) and Direcci\'{o}n General de Asuntos del Personal Academico (DGAPA), Mexico;
Nederlandse Organisatie voor Wetenschappelijk Onderzoek (NWO), Netherlands;
The Research Council of Norway, Norway;
Commission on Science and Technology for Sustainable Development in the South (COMSATS), Pakistan;
Pontificia Universidad Cat\'{o}lica del Per\'{u}, Peru;
Ministry of Science and Higher Education and National Science Centre, Poland;
Korea Institute of Science and Technology Information and National Research Foundation of Korea (NRF), Republic of Korea;
Ministry of Education and Scientific Research, Institute of Atomic Physics and Ministry of Research and Innovation and Institute of Atomic Physics, Romania;
Joint Institute for Nuclear Research (JINR), Ministry of Education and Science of the Russian Federation, National Research Centre Kurchatov Institute, Russian Science Foundation and Russian Foundation for Basic Research, Russia;
Ministry of Education, Science, Research and Sport of the Slovak Republic, Slovakia;
National Research Foundation of South Africa, South Africa;
Swedish Research Council (VR) and Knut \& Alice Wallenberg Foundation (KAW), Sweden;
European Organization for Nuclear Research, Switzerland;
National Science and Technology Development Agency (NSDTA), Suranaree University of Technology (SUT) and Office of the Higher Education Commission under NRU project of Thailand, Thailand;
Turkish Atomic Energy Agency (TAEK), Turkey;
National Academy of  Sciences of Ukraine, Ukraine;
Science and Technology Facilities Council (STFC), United Kingdom;
National Science Foundation of the United States of America (NSF) and United States Department of Energy, Office of Nuclear Physics (DOE NP), United States of America.
\end{acknowledgement}
\ifbibtex
\bibliographystyle{utphys}
\bibliography{biblio}{}
\else
\input{refpreprint.tex}
\fi
\newpage
\appendix
\section{The ALICE Collaboration}
\label{app:collab}

\begingroup
\small
\begin{flushleft}
S.~Acharya\Irefn{org141}\And 
D.~Adamov\'{a}\Irefn{org93}\And 
S.P.~Adhya\Irefn{org141}\And 
A.~Adler\Irefn{org74}\And 
J.~Adolfsson\Irefn{org80}\And 
M.M.~Aggarwal\Irefn{org98}\And 
G.~Aglieri Rinella\Irefn{org34}\And 
M.~Agnello\Irefn{org31}\And 
N.~Agrawal\Irefn{org10}\And 
Z.~Ahammed\Irefn{org141}
% ...
\And
\renewcommand\labelenumi{\textsuperscript{\theenumi}~}

\section*{Affiliation notes}
\renewcommand\theenumi{\roman{enumi}}
\begin{Authlist}
\item \Adef{org*}Deceased
\item \Adef{orgI}Dipartimento DET del Politecnico di Torino, Turin, Italy
\item \Adef{orgII}M.V. Lomonosov Moscow State University, D.V. Skobeltsyn Institute of Nuclear, Physics, Moscow, Russia
\item \Adef{orgIII}Department of Applied Physics, Aligarh Muslim University, Aligarh, India
\item \Adef{orgIV}Institute of Theoretical Physics, University of Wroclaw, Poland
\end{Authlist}

\section*{Collaboration Institutes}
\renewcommand\theenumi{\arabic{enumi}~}
\begin{Authlist}
\item \Idef{org1}A.I. Alikhanyan National Science Laboratory (Yerevan Physics Institute) Foundation, Yerevan, Armenia
\item \Idef{org2}Bogolyubov Institute for Theoretical Physics, National Academy of Sciences of Ukraine, Kiev, Ukraine
\item \Idef{org3}Bose Institute, Department of Physics  and Centre for Astroparticle Physics and Space Science (CAPSS), Kolkata, India
\item \Idef{org4}Budker Institute for Nuclear Physics, Novosibirsk, Russia
\item \Idef{org5}California Polytechnic State University, San Luis Obispo, California, United States
\item \Idef{org6}Central China Normal University, Wuhan, China
\item \Idef{org7}Centre de Calcul de l'IN2P3, Villeurbanne, Lyon, France
\item \Idef{org8}Centro de Aplicaciones Tecnol\'{o}gicas y Desarrollo Nuclear (CEADEN), Havana, Cuba
\item \Idef{org9}Centro de Investigaci\'{o}n y de Estudios Avanzados (CINVESTAV), Mexico City and M\'{e}rida, Mexico
\item \Idef{org10}Centro Fermi - Museo Storico della Fisica e Centro Studi e Ricerche ``Enrico Fermi', Rome, Italy
\item \Idef{org11}Chicago State University, Chicago, Illinois, United States
\item \Idef{org12}China Institute of Atomic Energy, Beijing, China
\item \Idef{org13}Chonbuk National University, Jeonju, Republic of Korea
\item \Idef{org14}Comenius University Bratislava, Faculty of Mathematics, Physics and Informatics, Bratislava, Slovakia
\item \Idef{org15}COMSATS University Islamabad, Islamabad, Pakistan
\item \Idef{org16}Creighton University, Omaha, Nebraska, United States
\item \Idef{org17}Department of Physics, Aligarh Muslim University, Aligarh, India
\item \Idef{org18}Department of Physics, Pusan National University, Pusan, Republic of Korea
\item \Idef{org19}Department of Physics, Sejong University, Seoul, Republic of Korea
\item \Idef{org20}Department of Physics, University of California, Berkeley, California, United States
\item \Idef{org21}Department of Physics, University of Oslo, Oslo, Norway
\item \Idef{org22}Department of Physics and Technology, University of Bergen, Bergen, Norway
\item \Idef{org23}Dipartimento di Fisica dell'Universit\`{a} 'La Sapienza' and Sezione INFN, Rome, Italy
\item \Idef{org24}Dipartimento di Fisica dell'Universit\`{a} and Sezione INFN, Cagliari, Italy
\item \Idef{org25}Dipartimento di Fisica dell'Universit\`{a} and Sezione INFN, Trieste, Italy
\item \Idef{org26}Dipartimento di Fisica dell'Universit\`{a} and Sezione INFN, Turin, Italy
\item \Idef{org27}Dipartimento di Fisica e Astronomia dell'Universit\`{a} and Sezione INFN, Bologna, Italy
\item \Idef{org28}Dipartimento di Fisica e Astronomia dell'Universit\`{a} and Sezione INFN, Catania, Italy
\item \Idef{org29}Dipartimento di Fisica e Astronomia dell'Universit\`{a} and Sezione INFN, Padova, Italy
\item \Idef{org30}Dipartimento di Fisica `E.R.~Caianiello' dell'Universit\`{a} and Gruppo Collegato INFN, Salerno, Italy
\item \Idef{org31}Dipartimento DISAT del Politecnico and Sezione INFN, Turin, Italy
\item \Idef{org32}Dipartimento di Scienze e Innovazione Tecnologica dell'Universit\`{a} del Piemonte Orientale and INFN Sezione di Torino, Alessandria, Italy
\item \Idef{org33}Dipartimento Interateneo di Fisica `M.~Merlin' and Sezione INFN, Bari, Italy
\item \Idef{org34}European Organization for Nuclear Research (CERN), Geneva, Switzerland
\item \Idef{org35}Faculty of Electrical Engineering, Mechanical Engineering and Naval Architecture, University of Split, Split, Croatia
\item \Idef{org36}Faculty of Engineering and Science, Western Norway University of Applied Sciences, Bergen, Norway
\item \Idef{org37}Faculty of Nuclear Sciences and Physical Engineering, Czech Technical University in Prague, Prague, Czech Republic
\item \Idef{org38}Faculty of Science, P.J.~\v{S}af\'{a}rik University, Ko\v{s}ice, Slovakia
\item \Idef{org39}Frankfurt Institute for Advanced Studies, Johann Wolfgang Goethe-Universit\"{a}t Frankfurt, Frankfurt, Germany
\item \Idef{org40}Gangneung-Wonju National University, Gangneung, Republic of Korea
\item \Idef{org41}Gauhati University, Department of Physics, Guwahati, India
\item \Idef{org42}Helmholtz-Institut f\"{u}r Strahlen- und Kernphysik, Rheinische Friedrich-Wilhelms-Universit\"{a}t Bonn, Bonn, Germany
\item \Idef{org43}Helsinki Institute of Physics (HIP), Helsinki, Finland
\item \Idef{org44}High Energy Physics Group,  Universidad Aut\'{o}noma de Puebla, Puebla, Mexico
\item \Idef{org45}Hiroshima University, Hiroshima, Japan
\item \Idef{org46}Hochschule Worms, Zentrum  f\"{u}r Technologietransfer und Telekommunikation (ZTT), Worms, Germany
\item \Idef{org47}Horia Hulubei National Institute of Physics and Nuclear Engineering, Bucharest, Romania
\item \Idef{org48}Indian Institute of Technology Bombay (IIT), Mumbai, India
\item \Idef{org49}Indian Institute of Technology Indore, Indore, India
\item \Idef{org50}Indonesian Institute of Sciences, Jakarta, Indonesia
\item \Idef{org51}INFN, Laboratori Nazionali di Frascati, Frascati, Italy
\item \Idef{org52}INFN, Sezione di Bari, Bari, Italy
\item \Idef{org53}INFN, Sezione di Bologna, Bologna, Italy
\item \Idef{org54}INFN, Sezione di Cagliari, Cagliari, Italy
\item \Idef{org55}INFN, Sezione di Catania, Catania, Italy
\item \Idef{org56}INFN, Sezione di Padova, Padova, Italy
\item \Idef{org57}INFN, Sezione di Roma, Rome, Italy
\item \Idef{org58}INFN, Sezione di Torino, Turin, Italy
\item \Idef{org59}INFN, Sezione di Trieste, Trieste, Italy
\item \Idef{org60}Inha University, Incheon, Republic of Korea
\item \Idef{org61}Institut de Physique Nucl\'{e}aire d'Orsay (IPNO), Institut National de Physique Nucl\'{e}aire et de Physique des Particules (IN2P3/CNRS), Universit\'{e} de Paris-Sud, Universit\'{e} Paris-Saclay, Orsay, France
\item \Idef{org62}Institute for Nuclear Research, Academy of Sciences, Moscow, Russia
\item \Idef{org63}Institute for Subatomic Physics, Utrecht University/Nikhef, Utrecht, Netherlands
\item \Idef{org64}Institute for Theoretical and Experimental Physics, Moscow, Russia
\item \Idef{org65}Institute of Experimental Physics, Slovak Academy of Sciences, Ko\v{s}ice, Slovakia
\item \Idef{org66}Institute of Physics, Homi Bhabha National Institute, Bhubaneswar, India
\item \Idef{org67}Institute of Physics of the Czech Academy of Sciences, Prague, Czech Republic
\item \Idef{org68}Institute of Space Science (ISS), Bucharest, Romania
\item \Idef{org69}Institut f\"{u}r Kernphysik, Johann Wolfgang Goethe-Universit\"{a}t Frankfurt, Frankfurt, Germany
\item \Idef{org70}Instituto de Ciencias Nucleares, Universidad Nacional Aut\'{o}noma de M\'{e}xico, Mexico City, Mexico
\item \Idef{org71}Instituto de F\'{i}sica, Universidade Federal do Rio Grande do Sul (UFRGS), Porto Alegre, Brazil
\item \Idef{org72}Instituto de F\'{\i}sica, Universidad Nacional Aut\'{o}noma de M\'{e}xico, Mexico City, Mexico
\item \Idef{org73}iThemba LABS, National Research Foundation, Somerset West, South Africa
\item \Idef{org74}Johann-Wolfgang-Goethe Universit\"{a}t Frankfurt Institut f\"{u}r Informatik, Fachbereich Informatik und Mathematik, Frankfurt, Germany
\item \Idef{org75}Joint Institute for Nuclear Research (JINR), Dubna, Russia
\item \Idef{org76}Korea Institute of Science and Technology Information, Daejeon, Republic of Korea
\item \Idef{org77}KTO Karatay University, Konya, Turkey
\item \Idef{org78}Laboratoire de Physique Subatomique et de Cosmologie, Universit\'{e} Grenoble-Alpes, CNRS-IN2P3, Grenoble, France
\item \Idef{org79}Lawrence Berkeley National Laboratory, Berkeley, California, United States
\item \Idef{org80}Lund University Department of Physics, Division of Particle Physics, Lund, Sweden
\item \Idef{org81}Nagasaki Institute of Applied Science, Nagasaki, Japan
\item \Idef{org82}Nara Women{'}s University (NWU), Nara, Japan
\item \Idef{org83}National and Kapodistrian University of Athens, School of Science, Department of Physics , Athens, Greece
\item \Idef{org84}National Centre for Nuclear Research, Warsaw, Poland
\item \Idef{org85}National Institute of Science Education and Research, Homi Bhabha National Institute, Jatni, India
\item \Idef{org86}National Nuclear Research Center, Baku, Azerbaijan
\item \Idef{org87}National Research Centre Kurchatov Institute, Moscow, Russia
\item \Idef{org88}Niels Bohr Institute, University of Copenhagen, Copenhagen, Denmark
\item \Idef{org89}Nikhef, National institute for subatomic physics, Amsterdam, Netherlands
\item \Idef{org90}NRC Kurchatov Institute IHEP, Protvino, Russia
\item \Idef{org91}NRNU Moscow Engineering Physics Institute, Moscow, Russia
\item \Idef{org92}Nuclear Physics Group, STFC Daresbury Laboratory, Daresbury, United Kingdom
\item \Idef{org93}Nuclear Physics Institute of the Czech Academy of Sciences, \v{R}e\v{z} u Prahy, Czech Republic
\item \Idef{org94}Oak Ridge National Laboratory, Oak Ridge, Tennessee, United States
\item \Idef{org95}Ohio State University, Columbus, Ohio, United States
\item \Idef{org96}Petersburg Nuclear Physics Institute, Gatchina, Russia
\item \Idef{org97}Physics department, Faculty of science, University of Zagreb, Zagreb, Croatia
\item \Idef{org98}Physics Department, Panjab University, Chandigarh, India
\item \Idef{org99}Physics Department, University of Jammu, Jammu, India
\item \Idef{org100}Physics Department, University of Rajasthan, Jaipur, India
\item \Idef{org101}Physikalisches Institut, Eberhard-Karls-Universit\"{a}t T\"{u}bingen, T\"{u}bingen, Germany
\item \Idef{org102}Physikalisches Institut, Ruprecht-Karls-Universit\"{a}t Heidelberg, Heidelberg, Germany
\item \Idef{org103}Physik Department, Technische Universit\"{a}t M\"{u}nchen, Munich, Germany
\item \Idef{org104}Politecnico di Bari, Bari, Italy
\item \Idef{org105}Research Division and ExtreMe Matter Institute EMMI, GSI Helmholtzzentrum f\"ur Schwerionenforschung GmbH, Darmstadt, Germany
\item \Idef{org106}Rudjer Bo\v{s}kovi\'{c} Institute, Zagreb, Croatia
\item \Idef{org107}Russian Federal Nuclear Center (VNIIEF), Sarov, Russia
\item \Idef{org108}Saha Institute of Nuclear Physics, Homi Bhabha National Institute, Kolkata, India
\item \Idef{org109}School of Physics and Astronomy, University of Birmingham, Birmingham, United Kingdom
\item \Idef{org110}Secci\'{o}n F\'{\i}sica, Departamento de Ciencias, Pontificia Universidad Cat\'{o}lica del Per\'{u}, Lima, Peru
\item \Idef{org111}Shanghai Institute of Applied Physics, Shanghai, China
\item \Idef{org112}St. Petersburg State University, St. Petersburg, Russia
\item \Idef{org113}Stefan Meyer Institut f\"{u}r Subatomare Physik (SMI), Vienna, Austria
\item \Idef{org114}SUBATECH, IMT Atlantique, Universit\'{e} de Nantes, CNRS-IN2P3, Nantes, France
\item \Idef{org115}Suranaree University of Technology, Nakhon Ratchasima, Thailand
\item \Idef{org116}Technical University of Ko\v{s}ice, Ko\v{s}ice, Slovakia
\item \Idef{org117}Technische Universit\"{a}t M\"{u}nchen, Excellence Cluster 'Universe', Munich, Germany
\item \Idef{org118}The Henryk Niewodniczanski Institute of Nuclear Physics, Polish Academy of Sciences, Cracow, Poland
\item \Idef{org119}The University of Texas at Austin, Austin, Texas, United States
\item \Idef{org120}Universidad Aut\'{o}noma de Sinaloa, Culiac\'{a}n, Mexico
\item \Idef{org121}Universidade de S\~{a}o Paulo (USP), S\~{a}o Paulo, Brazil
\item \Idef{org122}Universidade Estadual de Campinas (UNICAMP), Campinas, Brazil
\item \Idef{org123}Universidade Federal do ABC, Santo Andre, Brazil
\item \Idef{org124}University College of Southeast Norway, Tonsberg, Norway
\item \Idef{org125}University of Cape Town, Cape Town, South Africa
\item \Idef{org126}University of Houston, Houston, Texas, United States
\item \Idef{org127}University of Jyv\"{a}skyl\"{a}, Jyv\"{a}skyl\"{a}, Finland
\item \Idef{org128}University of Liverpool, Liverpool, United Kingdom
\item \Idef{org129}University of Science and Techonology of China, Hefei, China
\item \Idef{org130}University of Tennessee, Knoxville, Tennessee, United States
\item \Idef{org131}University of the Witwatersrand, Johannesburg, South Africa
\item \Idef{org132}University of Tokyo, Tokyo, Japan
\item \Idef{org133}University of Tsukuba, Tsukuba, Japan
\item \Idef{org134}Universit\'{e} Clermont Auvergne, CNRS/IN2P3, LPC, Clermont-Ferrand, France
\item \Idef{org135}Universit\'{e} de Lyon, Universit\'{e} Lyon 1, CNRS/IN2P3, IPN-Lyon, Villeurbanne, Lyon, France
\item \Idef{org136}Universit\'{e} de Strasbourg, CNRS, IPHC UMR 7178, F-67000 Strasbourg, France, Strasbourg, France
\item \Idef{org137}Universit\'{e} Paris-Saclay Centre d'Etudes de Saclay (CEA), IRFU, D\'{e}partment de Physique Nucl\'{e}aire (DPhN), Saclay, France
\item \Idef{org138}Universit\`{a} degli Studi di Foggia, Foggia, Italy
\item \Idef{org139}Universit\`{a} degli Studi di Pavia, Pavia, Italy
\item \Idef{org140}Universit\`{a} di Brescia, Brescia, Italy
\item \Idef{org141}Variable Energy Cyclotron Centre, Homi Bhabha National Institute, Kolkata, India
\item \Idef{org142}Warsaw University of Technology, Warsaw, Poland
\item \Idef{org143}Wayne State University, Detroit, Michigan, United States
\item \Idef{org144}Westf\"{a}lische Wilhelms-Universit\"{a}t M\"{u}nster, Institut f\"{u}r Kernphysik, M\"{u}nster, Germany
\item \Idef{org145}Wigner Research Centre for Physics, Hungarian Academy of Sciences, Budapest, Hungary
\item \Idef{org146}Yale University, New Haven, Connecticut, United States
\item \Idef{org147}Yonsei University, Seoul, Republic of Korea
\end{Authlist}
\endgroup
\else
\ifbibtex
\bibliographystyle{utphys}
\bibliography{biblio}{}
\else
\input{refpreprint.tex}
\fi
\fi
\else
\iffull

\input{refpaper.tex}
\else
\ifbibtex
\bibliographystyle{utphys}
\bibliography{biblio}{}

\providecommand{\href}[2]{#2}\begingroup\raggedright\begin{thebibliography}{10}

\bibitem{Bhattacharya:2014ara}
{\bfseries HotQCD} Collaboration, T.~Bhattacharya, M.~I. Buchoff, N.~H. Christ,
  et~al., ``{QCD Phase Transition with Chiral Quarks and Physical Quark
  Masses},'' {\em Phys. Rev. Lett.} {\bfseries 113} (Aug, 2014) 082001,
  \href{http://arxiv.org/abs/1402.5175}{{\ttfamily arXiv:1402.5175 [hep-lat]}}.

\bibitem{Arsene20051}
{\bfseries BRAHMS} Collaboration, I.~Arsene et~al., ``Quark--gluon plasma and
  color glass condensate at {RHIC}? {T}he perspective from the {BRAHMS}
  experiment,'' {\em Nucl.\ Phys.} {\bfseries A757} no.~1-2, (2005) 1--27,
  \href{http://arxiv.org/abs/nucl-ex/0410020}{{\ttfamily nucl-ex/0410020}}.

\bibitem{Back200528}
{\bfseries PHOBOS} Collaboration, B.~Back et~al., ``The {PHOBOS} perspective on
  discoveries at {RHIC},'' {\em Nucl.\ Phys.} {\bfseries A757} no.~1-2, (2005)
  28--101, \href{http://arxiv.org/abs/nucl-ex/0410003}{{\ttfamily
  nucl-ex/0410003}}.

\bibitem{Adcox2005184}
{\bfseries PHENIX} Collaboration, K.~Adcox et~al., ``Formation of dense
  partonic matter in relativistic nucleus--nucleus collisions at {RHIC}:
  {E}xperimental evaluation by the {PHENIX} collaboration,'' {\em Nucl.\ Phys.}
  {\bfseries A757} no.~1-2, (2005) 184--283,
  \href{http://arxiv.org/abs/nucl-ex/0410022}{{\ttfamily nucl-ex/0410022}}.

\bibitem{Adams2005102}
{\bfseries STAR} Collaboration, J.~Adams et~al., ``{Experimental and
  theoretical challenges in the search for the quark--gluon plasma: The STAR
  Collaboration's critical assessment of the evidence from RHIC collisions},''
  {\em Nucl.\ Phys.} {\bfseries A757} no.~1-2, (2005) 102--183,
  \href{http://arxiv.org/abs/nucl-ex/0501009}{{\ttfamily nucl-ex/0501009}}.

\bibitem{Aamodt:2010pb}
{\bfseries ALICE} Collaboration, K.~Aamodt et~al., ``{Charged-particle
  multiplicity density at mid-rapidity in central Pb--Pb collisions at
  $\snn=2.76$ {TeV}},'' {\em Phys.Rev.Lett.} {\bfseries 105} (2010) 252301,
\href{http://arxiv.org/abs/1011.3916}{{\ttfamily arXiv:1011.3916 [nucl-ex]}}.

\bibitem{Aamodt:2010cz}
{\bfseries ALICE} Collaboration, K.~Aamodt et~al., ``{Centrality dependence of
  the charged-particle multiplicity density at mid-rapidity in Pb--Pb
  collisions at $\snn=2.76$ TeV},''
  \href{http://dx.doi.org/10.1103/PhysRevLett.106.032301}{{\em Phys.Rev.Lett.}
  {\bfseries 106} (2011) 032301},
\href{http://arxiv.org/abs/1012.1657}{{\ttfamily arXiv:1012.1657 [nucl-ex]}}.

\bibitem{Chatrchyan:2011pb}
{\bfseries CMS} Collaboration, S.~Chatrchyan et~al., ``{Dependence on
  pseudorapidity and centrality of charged hadron production in Pb--Pb
  collisions at a nucleon-nucleon centre-of-mass energy of 2.76 TeV},''
  \href{http://dx.doi.org/10.1007/JHEP08(2011)141}{{\em JHEP} {\bfseries 1108}
  (2011) 141},
\href{http://arxiv.org/abs/1107.4800}{{\ttfamily arXiv:1107.4800 [nucl-ex]}}.

\bibitem{Aamodt:2011mr}
{\bfseries ALICE} Collaboration, K.~Aamodt et~al., ``{Two-pion Bose-Einstein
  correlations in central Pb--Pb collisions at $\snn=2.76$ TeV},''
  \href{http://dx.doi.org/10.1016/j.physletb.2010.12.053}{{\em Phys.Lett.}
  {\bfseries B696} (2011) 328--337},
\href{http://arxiv.org/abs/1012.4035}{{\ttfamily arXiv:1012.4035 [nucl-ex]}}.

\bibitem{Aamodt:2010pa}
{\bfseries ALICE} Collaboration, K.~Aamodt et~al., ``{Elliptic flow of charged
  particles in Pb--Pb collisions at 2.76 TeV},''
  \href{http://dx.doi.org/10.1103/PhysRevLett.105.252302}{{\em Phys.Rev.Lett.}
  {\bfseries 105} (2010) 252302},
\href{http://arxiv.org/abs/1011.3914}{{\ttfamily arXiv:1011.3914 [nucl-ex]}}.

\bibitem{ATLAS:2011ah}
{\bfseries ATLAS} Collaboration, G.~Aad et~al., ``{Measurement of the
  pseudorapidity and transverse momentum dependence of the elliptic flow of
  charged particles in lead-lead collisions at $\snn=2.76$ TeV with the ATLAS
  detector},'' \href{http://dx.doi.org/10.1016/j.physletb.2011.12.056}{{\em
  Phys.Lett.} {\bfseries B707} (2012) 330--348},
\href{http://arxiv.org/abs/1108.6018}{{\ttfamily arXiv:1108.6018 [hep-ex]}}.

\bibitem{Chatrchyan:2012wg}
{\bfseries CMS} Collaboration, S.~Chatrchyan et~al., ``{Centrality dependence
  of dihadron correlations and azimuthal anisotropy harmonics in Pb--Pb
  collisions at $\snn=2.76$ TeV},''
  \href{http://dx.doi.org/10.1140/epjc/s10052-012-2012-3}{{\em Eur.Phys.J.}
  {\bfseries C72} (2012) 2012},
\href{http://arxiv.org/abs/1201.3158}{{\ttfamily arXiv:1201.3158 [nucl-ex]}}.

\bibitem{ALICE:2011ab}
{\bfseries ALICE} Collaboration, K.~Aamodt et~al., ``{Higher harmonic
  anisotropic flow measurements of charged particles in Pb--Pb collisions at
  $\snn=2.76$ {TeV}},''
  \href{http://dx.doi.org/10.1103/PhysRevLett.107.032301}{{\em Phys.Rev.Lett.}
  {\bfseries 107} (2011) 032301},
\href{http://arxiv.org/abs/1105.3865}{{\ttfamily arXiv:1105.3865 [nucl-ex]}}.

\bibitem{Aad:2013xma}
{\bfseries ATLAS} Collaboration, G.~Aad et~al., ``{Measurement of the
  distributions of event-by-event flow harmonics in lead-lead collisions at
  2.76 TeV with the ATLAS detector at the LHC},''
  \href{http://dx.doi.org/10.1007/JHEP11(2013)183}{{\em JHEP} {\bfseries 1311}
  (2013) 183},
\href{http://arxiv.org/abs/1305.2942}{{\ttfamily arXiv:1305.2942 [hep-ex]}}.

\bibitem{Chatrchyan:2013kba}
{\bfseries CMS} Collaboration, S.~Chatrchyan et~al., ``{Measurement of
  higher-order harmonic azimuthal anisotropy in Pb--Pb collisions at
  $\snn=2.76$ TeV},'' \href{http://dx.doi.org/10.1103/PhysRevC.89.044906}{{\em
  Phys.Rev.} {\bfseries C89} (2014) 044906},
\href{http://arxiv.org/abs/1310.8651}{{\ttfamily arXiv:1310.8651 [nucl-ex]}}.

\bibitem{Aamodt:2010jd}
{\bfseries ALICE} Collaboration, K.~Aamodt et~al., ``{Suppression of charged
  particle Production at Large Transverse Momentum in Central Pb--Pb Collisions
  at $\snn=2.76$ TeV},''
  \href{http://dx.doi.org/10.1016/j.physletb.2010.12.020}{{\em Phys.Lett.}
  {\bfseries B696} (2011) 30--39},
\href{http://arxiv.org/abs/1012.1004}{{\ttfamily arXiv:1012.1004 [nucl-ex]}}.

\bibitem{Chatrchyan:2011sx}
{\bfseries CMS} Collaboration, S.~Chatrchyan et~al., ``{Observation and studies
  of jet quenching in Pb--Pb collisions at nucleon-nucleon center-of-mass
  energy of 2.76 TeV},''
  \href{http://dx.doi.org/10.1103/PhysRevC.84.024906}{{\em Phys.Rev.}
  {\bfseries C84} (2011) 024906},
\href{http://arxiv.org/abs/1102.1957}{{\ttfamily arXiv:1102.1957 [nucl-ex]}}.

\bibitem{Gyulassy:1990ye}
M.~Gyulassy and M.~Plumer, ``{Jet Quenching in Dense Matter},''
{\em Phys.Lett.} {\bfseries B243} (1990) 432--438.

\bibitem{Baier:1994bd}
R.~Baier, Y.~L. Dokshitzer, S.~Peigne, and D.~Schiff, ``{{Induced gluon
  radiation in a QCD medium}},''
  \href{http://dx.doi.org/10.1016/0370-2693(94)01617-L}{{\em Phys.Lett.}
  {\bfseries B345} (1995) 277--286},
\href{http://arxiv.org/abs/hep-ph/9411409}{{\ttfamily arXiv:hep-ph/9411409
  [hep-ph]}}.

\bibitem{Salgado:2003rv}
C.~A. Salgado and U.~A. Wiedemann, ``{Medium modification of jet shapes and jet
  multiplicities},''
  \href{http://dx.doi.org/10.1103/PhysRevLett.93.042301}{{\em Phys.Rev.Lett.}
  {\bfseries 93} (2004) 042301},
\href{http://arxiv.org/abs/hep-ph/0310079}{{\ttfamily arXiv:hep-ph/0310079
  [hep-ph]}}.

\bibitem{Adcox:2001jp}
{\bfseries PHENIX} Collaboration, K.~Adcox et~al., ``{Suppression of hadrons
  with large transverse momentum in central Au+Au collisions at $\snn=130$
  GeV},'' \href{http://dx.doi.org/10.1103/PhysRevLett.88.022301}{{\em
  Phys.Rev.Lett.} {\bfseries 88} (2002) 022301},
\href{http://arxiv.org/abs/nucl-ex/0109003}{{\ttfamily arXiv:nucl-ex/0109003
  [nucl-ex]}}.

\bibitem{Adler:2002tq}
{\bfseries STAR} Collaboration, C.~Adler et~al., ``{Disappearance of
  back-to-back high $\pT$ hadron correlations in central Au+Au collisions at
  $\snn=200$ GeV},''
  \href{http://dx.doi.org/10.1103/PhysRevLett.90.082302}{{\em Phys.Rev.Lett.}
  {\bfseries 90} (2003) 082302},
\href{http://arxiv.org/abs/nucl-ex/0210033}{{\ttfamily arXiv:nucl-ex/0210033
  [nucl-ex]}}.

\bibitem{Adler:2002xw}
{\bfseries STAR} Collaboration, C.~Adler et~al., ``{Centrality dependence of
  high $p_{T}$ hadron suppression in Au+Au collisions at $\snn=130$~GeV},''
  \href{http://dx.doi.org/10.1103/PhysRevLett.89.202301}{{\em Phys. Rev. Lett.}
  {\bfseries 89} (2002) 202301},
\href{http://arxiv.org/abs/nucl-ex/0206011}{{\ttfamily arXiv:nucl-ex/0206011
  [nucl-ex]}}.

\bibitem{Adcox:2002pe}
{\bfseries PHENIX} Collaboration, K.~Adcox et~al., ``{Centrality dependence of
  the high-\pT\ charged hadron suppression in Au+Au collisions at $\snn=130$
  GeV},'' \href{http://dx.doi.org/10.1016/S0370-2693(03)00423-4}{{\em
  Phys.Lett.} {\bfseries B561} (2003) 82--92},
\href{http://arxiv.org/abs/nucl-ex/0207009}{{\ttfamily arXiv:nucl-ex/0207009
  [nucl-ex]}}.

\bibitem{Adler:2003qi}
{\bfseries PHENIX} Collaboration, S.~S. Adler et~al., ``{Suppressed $\pi^0$
  production at large transverse momentum in central Au+Au collisions at
  $\snn=200$ GeV},''
  \href{http://dx.doi.org/10.1103/PhysRevLett.91.072301}{{\em Phys.Rev.Lett.}
  {\bfseries 91} (2003) 072301},
\href{http://arxiv.org/abs/nucl-ex/0304022}{{\ttfamily arXiv:nucl-ex/0304022}}.

\bibitem{Adams:2003kv}
{\bfseries STAR} Collaboration, J.~Adams et~al., ``{Transverse-momentum and
  collision-energy dependence of high-\pT\ hadron suppression in Au+Au
  collisions at ultrarelativistic energies},''
  \href{http://dx.doi.org/10.1103/PhysRevLett.91.172302}{{\em Phys. Rev. Lett.}
  {\bfseries 91} (Oct, 2003) 172302},
  \href{http://arxiv.org/abs/nucl-ex/0305015}{{\ttfamily arXiv:nucl-ex/0305015
  [nucl-ex]}}.

\bibitem{Adams:2003im}
{\bfseries STAR} Collaboration, J.~Adams et~al., ``{Evidence from d+Au
  measurements for final state suppression of high-\pT\ hadrons in Au+Au
  collisions at RHIC},''
  \href{http://dx.doi.org/10.1103/PhysRevLett.91.072304}{{\em Phys.Rev.Lett.}
  {\bfseries 91} (2003) 072304},
\href{http://arxiv.org/abs/nucl-ex/0306024}{{\ttfamily arXiv:nucl-ex/0306024
  [nucl-ex]}}.

\bibitem{Back:2003qr}
{\bfseries PHOBOS} Collaboration, B.~Back et~al., ``{Charged hadron transverse
  momentum distributions in Au+Au collisions at $\snn=200$ GeV},''
  \href{http://dx.doi.org/10.1016/j.physletb.2003.10.101}{{\em Phys.Lett.}
  {\bfseries B578} (2004) 297--303},
\href{http://arxiv.org/abs/nucl-ex/0302015}{{\ttfamily arXiv:nucl-ex/0302015
  [nucl-ex]}}.

\bibitem{Arsene:2003yk}
{\bfseries BRAHMS} Collaboration, I.~Arsene et~al., ``{Transverse momentum
  spectra in Au+Au and d+Au collisions at $\snn=200$ GeV and the pseudorapidity
  dependence of high-\pT\ suppression},''
  \href{http://dx.doi.org/10.1103/PhysRevLett.91.072305}{{\em Phys.Rev.Lett.}
  {\bfseries 91} (2003) 072305},
\href{http://arxiv.org/abs/nucl-ex/0307003}{{\ttfamily arXiv:nucl-ex/0307003
  [nucl-ex]}}.

\bibitem{Adams:2006yt}
{\bfseries STAR} Collaboration, J.~Adams et~al., ``{Direct observation of
  dijets in central Au--Au collisions at $\snn=200$ GeV},''
  \href{http://dx.doi.org/10.1103/PhysRevLett.97.162301}{{\em Phys. Rev. Lett.}
  {\bfseries 97} (2006) 162301},
\href{http://arxiv.org/abs/nucl-ex/0604018}{{\ttfamily arXiv:nucl-ex/0604018
  [nucl-ex]}}.

\bibitem{Adare:2006nr}
{\bfseries PHENIX} Collaboration, A.~Adare et~al., ``{System Size and Energy
  Dependence of Jet-Induced Hadron Pair Correlation Shapes in Cu+Cu and Au+Au
  Collisions at $\snn=200$ and 62.4 GeV},''
  \href{http://dx.doi.org/10.1103/PhysRevLett.98.232302}{{\em Phys.Rev.Lett.}
  {\bfseries 98} (2007) 232302},
\href{http://arxiv.org/abs/nucl-ex/0611019}{{\ttfamily arXiv:nucl-ex/0611019
  [nucl-ex]}}.

\bibitem{Adare:2008cg}
{\bfseries PHENIX} Collaboration, A.~Adare et~al., ``{Quantitative constraints
  on the opacity of hot partonic matter from semi-inclusive single high
  transverse momentum pion suppression in Au+Au collisions at $\snn=200$
  GeV},'' \href{http://dx.doi.org/10.1103/PhysRevC.77.064907}{{\em Phys. Rev.}
  {\bfseries C77} (2008) 064907},
\href{http://arxiv.org/abs/0801.1665}{{\ttfamily arXiv:0801.1665 [nucl-ex]}}.

\bibitem{Adamczyk:2013jei}
{\bfseries STAR} Collaboration, L.~Adamczyk et~al., ``{Jet-Hadron Correlations
  in $\sqrt{s_{NN}} = 200$ GeV $p+p$ and Central $Au+Au$ Collisions},''
  \href{http://dx.doi.org/10.1103/PhysRevLett.112.122301}{{\em Phys. Rev.
  Lett.} {\bfseries 112} no.~12, (2014) 122301},
\href{http://arxiv.org/abs/1302.6184}{{\ttfamily arXiv:1302.6184 [nucl-ex]}}.

\bibitem{Adamczyk:2017yhe}
{\bfseries STAR} Collaboration, L.~Adamczyk et~al., ``{Measurements of jet
  quenching with semi-inclusive hadron-jet distributions in Au--Au collisions
  at $\snn$ = 200 GeV},''
  \href{http://dx.doi.org/10.1103/PhysRevC.96.024905}{{\em Phys. Rev.}
  {\bfseries C96} no.~2, (2017) 024905},
\href{http://arxiv.org/abs/1702.01108}{{\ttfamily arXiv:1702.01108 [nucl-ex]}}.

\bibitem{Aad:2010bu}
{\bfseries ATLAS} Collaboration, G.~Aad et~al., ``{Observation of a
  centrality-dependent dijet asymmetry in Pb--Pb collisions at $\snn=2.76$ TeV
  with the ATLAS Detector at the LHC},''
  \href{http://dx.doi.org/10.1103/PhysRevLett.105.252303}{{\em Phys.Rev.Lett.}
  {\bfseries 105} (2010) 252303},
\href{http://arxiv.org/abs/1011.6182}{{\ttfamily arXiv:1011.6182 [hep-ex]}}.

\bibitem{Aamodt:2011vg}
{\bfseries ALICE} Collaboration, K.~Aamodt et~al., ``{Particle-yield
  modification in jet-like azimuthal di-hadron correlations in Pb--Pb
  collisions at $\snn=2.76$ TeV},''
  \href{http://dx.doi.org/10.1103/PhysRevLett.108.092301}{{\em Phys.Rev.Lett.}
  {\bfseries 108} (2012) 092301},
\href{http://arxiv.org/abs/1110.0121}{{\ttfamily arXiv:1110.0121 [nucl-ex]}}.

\bibitem{CMS:2012aa}
{\bfseries CMS} Collaboration, S.~Chatrchyan et~al., ``{Study of high-\pT\
  charged particle suppression in Pb--Pb compared to $pp$ collisions at
  $\snn=2.76$ TeV},''
  \href{http://dx.doi.org/10.1140/epjc/s10052-012-1945-x}{{\em Eur.Phys.J.}
  {\bfseries C72} (2012) 1945},
\href{http://arxiv.org/abs/1202.2554}{{\ttfamily arXiv:1202.2554 [nucl-ex]}}.

\bibitem{Chatrchyan:2012nia}
{\bfseries CMS} Collaboration, S.~Chatrchyan et~al., ``{Jet momentum dependence
  of jet quenching in Pb--Pb collisions at $\snn=2.76$ TeV},''
  \href{http://dx.doi.org/10.1016/j.physletb.2012.04.058}{{\em Phys.Lett.}
  {\bfseries B712} (2012) 176--197},
\href{http://arxiv.org/abs/1202.5022}{{\ttfamily arXiv:1202.5022 [nucl-ex]}}.

\bibitem{Chatrchyan:2012gw}
{\bfseries CMS} Collaboration, S.~Chatrchyan et~al., ``{Measurement of jet
  fragmentation into charged particles in $pp$ and Pb--Pb collisions at
  $\snn=2.76$ TeV},'' \href{http://dx.doi.org/10.1007/JHEP10(2012)087}{{\em
  JHEP} {\bfseries 1210} (2012) 087},
\href{http://arxiv.org/abs/1205.5872}{{\ttfamily arXiv:1205.5872 [nucl-ex]}}.

\bibitem{Chatrchyan:2012gt}
{\bfseries CMS} Collaboration, S.~Chatrchyan et~al., ``{Studies of jet
  quenching using isolated-photon+jet correlations in Pb--Pb and $pp$
  collisions at $\snn=2.76$ TeV},''
  \href{http://dx.doi.org/10.1016/j.physletb.2012.11.003}{{\em Phys.Lett.}
  {\bfseries B718} (2013) 773--794},
\href{http://arxiv.org/abs/1205.0206}{{\ttfamily arXiv:1205.0206 [nucl-ex]}}.

\bibitem{Aad:2012vca}
{\bfseries ATLAS} Collaboration, G.~Aad et~al., ``{Measurement of the jet
  radius and transverse momentum dependence of inclusive jet suppression in
  lead-lead collisions at $\snn=2.76$ TeV with the ATLAS detector},''
  \href{http://dx.doi.org/10.1016/j.physletb.2013.01.024}{{\em Phys.Lett.}
  {\bfseries B719} (2013) 220--241},
\href{http://arxiv.org/abs/1208.1967}{{\ttfamily arXiv:1208.1967 [hep-ex]}}.

\bibitem{Chatrchyan:2013exa}
{\bfseries CMS} Collaboration, S.~Chatrchyan et~al., ``{Evidence of b-jet
  quenching in Pb--Pb collisions at $\snn=2.76$ TeV},''
  \href{http://dx.doi.org/10.1103/PhysRevLett.113.132301}{{\em Phys.Rev.Lett.}
  {\bfseries 113} no.~13, (2014) 132301},
\href{http://arxiv.org/abs/1312.4198}{{\ttfamily arXiv:1312.4198 [nucl-ex]}}.

\bibitem{Chatrchyan:2013kwa}
{\bfseries CMS} Collaboration, S.~Chatrchyan et~al., ``{Modification of jet
  shapes in Pb--Pb collisions at $\snn=2.76$ TeV},''
  \href{http://dx.doi.org/10.1016/j.physletb.2014.01.042}{{\em Phys.Lett.}
  {\bfseries B730} (2014) 243--263},
\href{http://arxiv.org/abs/1310.0878}{{\ttfamily arXiv:1310.0878 [nucl-ex]}}.

\bibitem{Chatrchyan:2014ava}
{\bfseries CMS} Collaboration, S.~Chatrchyan et~al., ``{Measurement of jet
  fragmentation in Pb--Pb and pp collisions at $\snn=2.76$ TeV},''
  \href{http://dx.doi.org/10.1103/PhysRevC.90.024908}{{\em Phys. Rev.}
  {\bfseries C90} no.~2, (2014) 024908},
\href{http://arxiv.org/abs/1406.0932}{{\ttfamily arXiv:1406.0932 [nucl-ex]}}.

\bibitem{Aad:2014wha}
{\bfseries ATLAS} Collaboration, G.~Aad et~al., ``{Measurement of inclusive jet
  charged-particle fragmentation functions in \PbPb\ collisions at $\snn=2.76$
  TeV with the ATLAS detector},''
  \href{http://dx.doi.org/10.1016/j.physletb.2014.10.065}{{\em Phys. Lett.}
  {\bfseries B739} (2014) 320--342},
\href{http://arxiv.org/abs/1406.2979}{{\ttfamily arXiv:1406.2979 [hep-ex]}}.

\bibitem{Aad:2014bxa}
{\bfseries ATLAS} Collaboration, G.~Aad et~al., ``{Measurements of the nuclear
  modification factor for jets in \PbPb\ collisions at $\snn=2.76$ TeV with the
  ATLAS Detector},''
  \href{http://dx.doi.org/10.1103/PhysRevLett.114.072302}{{\em Phys. Rev.
  Lett.} {\bfseries 114} no.~7, (2015) 072302},
\href{http://arxiv.org/abs/1411.2357}{{\ttfamily arXiv:1411.2357 [hep-ex]}}.

\bibitem{Adam:2015doa}
{\bfseries ALICE} Collaboration, J.~Adam et~al., ``{Measurement of jet
  quenching with semi-inclusive hadron-jet distributions in central \PbPb\
  collisions at $\snn=2.76$ TeV},''
  \href{http://dx.doi.org/10.1007/JHEP09(2015)170}{{\em JHEP} {\bfseries 09}
  (2015) 170},
\href{http://arxiv.org/abs/1506.03984}{{\ttfamily arXiv:1506.03984 [nucl-ex]}}.

\bibitem{Adam:2016a}
{\bfseries ALICE} Collaboration, J.~Adam et~al., ``{Anomalous evolution of the
  near-side jet peak shape in Pb--Pb collisions at $\snn=2.76$ TeV},''
  \href{http://dx.doi.org/10.1103/PhysRevLett.119.102301}{{\em Phys.Rev.Lett.}
  {\bfseries 119} (2017) 102301},
  \href{http://arxiv.org/abs/1609.06643}{{\ttfamily arXiv:1609.06643
  [nucl-ex]}}.

\bibitem{Adam:2016b}
{\bfseries ALICE} Collaboration, J.~Adam et~al., ``{Evolution of the
  longitudinal and azimuthal structure of the near-side jet peak in Pb--Pb
  collisions at $\snn=2.76$ TeV},''
  \href{http://dx.doi.org/10.1103/PhysRevC.96.034904}{{\em Phys.Rev.C.}
  {\bfseries 96} (2017) 034904},
  \href{http://arxiv.org/abs/1609.06667}{{\ttfamily arXiv:1609.06667
  [nucl-ex]}}.

\bibitem{CMS:2016a}
{\bfseries CMS} Collaboration, ``{Correlations between jets and charged
  particles in PbPb and pp collisions at $\snn=2.76$ TeV},''
  \href{http://dx.doi.org/10.1007/JHEP02(2016)156}{{\em JHEP} {\bfseries 1602}
  (2016) 156}, \href{http://arxiv.org/abs/1601.00079}{{\ttfamily
  arXiv:1601.00079 [nucl-ex]}}.

\bibitem{Acharya:2018uvf}
{\bfseries ALICE} Collaboration, S.~Acharya et~al., ``{Medium modification of
  the shape of small-radius jets in central Pb--Pb collisions at $\snn =
  2.76\,\rm{TeV}$},'' \href{http://dx.doi.org/10.1007/JHEP10(2018)139}{{\em
  JHEP} {\bfseries 10} (2018) 139},
\href{http://arxiv.org/abs/1807.06854}{{\ttfamily arXiv:1807.06854 [nucl-ex]}}.

\bibitem{Sjostrand:2006za}
T.~Sjostrand, S.~Mrenna, and P.~Z. Skands, ``{PYTHIA 6.4 Physics and Manual},''
  \href{http://dx.doi.org/10.1088/1126-6708/2006/05/026}{{\em JHEP} {\bfseries
  0605} (2006) 026},
\href{http://arxiv.org/abs/hep-ph/0603175}{{\ttfamily arXiv:hep-ph/0603175
  [hep-ph]}}.

\bibitem{Aamodt:2008zz}
{\bfseries ALICE} Collaboration, K.~Aamodt et~al., ``{The ALICE experiment at
  the CERN LHC},''
\href{http://dx.doi.org/10.1088/1748-0221/3/08/S08002}{{\em JINST} {\bfseries
  3} (2008) S08002}.

\bibitem{Abelev:2014ffa}
{\bfseries ALICE} Collaboration, B.~B. Abelev et~al., ``{Performance of the
  ALICE Experiment at the CERN LHC},''
  \href{http://dx.doi.org/10.1142/S0217751X14300440}{{\em Int.J.Mod.Phys.}
  {\bfseries A29} (2014) 1430044},
\href{http://arxiv.org/abs/1402.4476}{{\ttfamily arXiv:1402.4476 [nucl-ex]}}.

\bibitem{Abelev:2013qoq}
{\bfseries ALICE} Collaboration, B.~Abelev et~al., ``{Centrality determination
  of Pb--Pb collisions at $\snn=2.76$ TeV with ALICE},''
  \href{http://dx.doi.org/10.1103/PhysRevC.88.044909}{{\em Phys.Rev.}
  {\bfseries C88} no.~4, (2013) 044909},
\href{http://arxiv.org/abs/1301.4361}{{\ttfamily arXiv:1301.4361 [nucl-ex]}}.

\bibitem{Abelev:2012hxa}
{\bfseries ALICE} Collaboration, B.~Abelev et~al., ``{Centrality Dependence of
  Charged Particle Production at Large Transverse Momentum in Pb--Pb Collisions
  at $\snn=2.76$ TeV},''
  \href{http://dx.doi.org/10.1016/j.physletb.2013.01.051}{{\em Phys.Lett.}
  {\bfseries B720} (2013) 52--62},
\href{http://arxiv.org/abs/1208.2711}{{\ttfamily arXiv:1208.2711 [hep-ex]}}.

\bibitem{hijing}
{X.-N. Wang, and M.\ Gyulassy}, ``{HIJING: A Monte Carlo model for multiple jet
  production in pp, pA and AA collisions},'' {\em Phys.\ Rev.\ D} {\bfseries
  44} (1991) 3501.

\bibitem{geant3ref2}
R.~Brun, F.~Carminati, and S.~Giani, ``{GEANT Detector Description and
  Simulation Tool},''
{\em CERN Program Library Long Write-up, W5013} (1994) .

\bibitem{ALICE-PUBLIC-2017-005}
{\bfseries ALICE} Collaboration, ``{The ALICE definition of primary
  particles},'' {\em ALICE-PUBLIC-2017-005} (Jun, 2017) .
  \url{https://cds.cern.ch/record/2270008}.

\bibitem{ABELEV:2013wsa}
{\bfseries ALICE} Collaboration, B.~Abelev et~al., ``{Long-range angular
  correlations of $\rm \pi$, K and p in p-Pb collisions at $\sqrt{s_{\rm NN}}$
  = 5.02 TeV},'' \href{http://dx.doi.org/10.1016/j.physletb.2013.08.024}{{\em
  Phys. Lett.} {\bfseries B726} (2013) 164--177},
\href{http://arxiv.org/abs/1307.3237}{{\ttfamily arXiv:1307.3237 [nucl-ex]}}.

\bibitem{Adam:2016ows}
{\bfseries ALICE} Collaboration, J.~Adam et~al., ``{Pseudorapidity dependence
  of the anisotropic flow of charged particles in Pb-Pb collisions at
  $\sqrt{s_{\rm NN}}=2.76$ TeV},''
  \href{http://dx.doi.org/10.1016/j.physletb.2016.07.017}{{\em Phys. Lett.}
  {\bfseries B762} (2016) 376--388},
\href{http://arxiv.org/abs/1605.02035}{{\ttfamily arXiv:1605.02035 [nucl-ex]}}.

\bibitem{Cacciari:2010te}
M.~Cacciari, J.~Rojo, G.~P. Salam, and G.~Soyez, ``{Jet Reconstruction in Heavy
  Ion Collisions},''
  \href{http://dx.doi.org/10.1140/epjc/s10052-011-1539-z}{{\em Eur.Phys.J.}
  {\bfseries C71} (2011) 1539},
\href{http://arxiv.org/abs/1010.1759}{{\ttfamily arXiv:1010.1759 [hep-ph]}}.

\bibitem{Cacciari:2008gp}
M.~Cacciari, G.~P. Salam, and G.~Soyez, ``{The Anti-k(t) jet clustering
  algorithm},'' \href{http://dx.doi.org/10.1088/1126-6708/2008/04/063}{{\em
  JHEP} {\bfseries 0804} (2008) 063},
\href{http://arxiv.org/abs/0802.1189}{{\ttfamily arXiv:0802.1189 [hep-ph]}}.

\bibitem{Cacciari:2011ma}
M.~Cacciari, G.~P. Salam, and G.~Soyez, ``{FastJet User Manual},''
  \href{http://dx.doi.org/10.1140/epjc/s10052-012-1896-2}{{\em Eur.Phys.J.}
  {\bfseries C72} (2012) 1896},
\href{http://arxiv.org/abs/1111.6097}{{\ttfamily arXiv:1111.6097 [hep-ph]}}.

\bibitem{Abelev:2012ej}
{\bfseries ALICE} Collaboration, B.~Abelev et~al., ``{Measurement of Event
  Background Fluctuations for Charged Particle Jet Reconstruction in Pb--Pb
  collisions at $\snn=2.76$ TeV},''
  \href{http://dx.doi.org/10.1007/JHEP03(2012)053}{{\em JHEP} {\bfseries 1203}
  (2012) 053},
\href{http://arxiv.org/abs/1201.2423}{{\ttfamily arXiv:1201.2423 [hep-ex]}}.

\bibitem{Abelev:2013kqa}
{\bfseries ALICE} Collaboration, B.~Abelev et~al., ``{Measurement of charged
  jet suppression in Pb--Pb collisions at $\snn=2.76$ TeV},''
  \href{http://dx.doi.org/10.1007/JHEP03(2014)013}{{\em JHEP} {\bfseries 1403}
  (2014) 013},
\href{http://arxiv.org/abs/1311.0633}{{\ttfamily arXiv:1311.0633 [nucl-ex]}}.

\bibitem{ALICE-PUBLIC-2019-002}
{\bfseries ALICE} Collaboration, ``{Supplemental figures for measurement of jet
  radial profiles in \PbPb\ collisions at $\sqrt{s_{NN}} = 2.76$~TeV},'' {\em
  ALICE-PUBLIC-2019-002} (2019) . \url{https://cds.cern.ch/record/2672661}.

\bibitem{Skands:2010ak}
P.~Z. Skands, ``{Tuning Monte Carlo Generators: The Perugia Tunes},''
  \href{http://dx.doi.org/10.1103/PhysRevD.82.074018}{{\em Phys.Rev.}
  {\bfseries D82} (2010) 074018},
\href{http://arxiv.org/abs/1005.3457}{{\ttfamily arXiv:1005.3457 [hep-ph]}}.

\bibitem{Spousta:2015fca}
M.~Spousta and B.~Cole, ``{Interpreting single jet measurements in Pb $+$ Pb
  collisions at the LHC},''
  \href{http://dx.doi.org/10.1140/epjc/s10052-016-3896-0}{{\em Eur. Phys. J.}
  {\bfseries C76} no.~2, (2016) 50},
\href{http://arxiv.org/abs/1504.05169}{{\ttfamily arXiv:1504.05169 [hep-ph]}}.

\bibitem{CasalderreySolana:2010eh}
J.~Casalderrey-Solana, J.~G. Milhano, and U.~A. Wiedemann, ``{Jet Quenching via
  Jet Collimation},''
  \href{http://dx.doi.org/10.1088/0954-3899/38/3/035006}{{\em J. Phys.}
  {\bfseries G38} (2011) 035006},
\href{http://arxiv.org/abs/1012.0745}{{\ttfamily arXiv:1012.0745 [hep-ph]}}.

\bibitem{CasalderreySolana:2012ef}
J.~Casalderrey-Solana, Y.~Mehtar-Tani, C.~A. Salgado, and K.~Tywoniuk, ``{New
  picture of jet quenching dictated by color coherence},''
  \href{http://dx.doi.org/10.1016/j.physletb.2013.07.046}{{\em Phys. Lett.}
  {\bfseries B725} (2013) 357--360},
\href{http://arxiv.org/abs/1210.7765}{{\ttfamily arXiv:1210.7765 [hep-ph]}}.

\end{thebibliography}\endgroup
\else
\input{refpaper.tex}
\fi
\fi
\fi
\end{document}